\documentclass[%
 reprint,
 superscriptaddress,
 amsmath,amssymb,
 aps,
 prx,
floatfix,
]{revtex4-2}

\usepackage{color}
\usepackage{graphicx}
\usepackage{dcolumn}
\usepackage{bm}
\usepackage{mathrsfs}
\usepackage{xr}
\usepackage{textcomp, gensymb}
\usepackage{float}

\newcommand{\Fig}[1]{Fig.~\ref{#1}}
\newcommand{\Eq}[1]{Eq.~\eqref{#1}}
\newcommand{\Appendix}[1]{Appendix~\ref{#1}}

\newcommand{\comment}[1]{}

\DeclareMathOperator*{\argmin}{arg\,min}

\newcommand{\appsection}[1]{\section{\MakeUppercase{#1}}}

\begin{document}

\preprint{APS/123-QED}
\title{A ``morphogenetic action'' principle for 3D shape formation by the growth of thin sheets}

\author{Dillon J. Cislo}
 \email{dilloncislo@gmail.com}
\affiliation{%
Department of Physics, University of California, Santa Barbara,  CA 93106, USA\looseness=-1}
\affiliation{%
Center for
Studies in Physics and Biology, Rockefeller University, New York, NY 10065, USA}
\author{Anastasios Pavlopoulos}
\affiliation{%
Institute of Molecular Biology and Biotechnology, Foundation for Research and Technology Hellas, 70013 Heraklion, Crete, Greece}
\author{Boris I. Shraiman}
 \email{shraiman@kitp.ucsb.edu}
\affiliation{%
Department of Physics, University of California, Santa Barbara,  CA 93106, USA\looseness=-1}
\affiliation{%
Kavli Institute for Theoretical Physics, University of California, Santa Barbara, CA 93106, USA}

\date{\today}

\begin{abstract}
How does growth encode form in developing organisms?
Many different spatiotemporal growth profiles may sculpt tissues into the same target 3D shapes, but only specific growth patterns are observed in animal and plant development.
In particular, growth profiles may differ in their degree of spatial variation and growth anisotropy, however, the criteria that distinguish observed patterns of growth from other possible alternatives are not understood.
Here we exploit the mathematical formalism of quasiconformal transformations to formulate the problem of ``growth pattern selection'' quantitatively in the context of 3D shape formation by growing 2D epithelial sheets.
We propose that nature settles on growth patterns that are the `simplest' in a certain way.
Specifically, we demonstrate that growth pattern selection can be formulated as an optimization problem and solved for the trajectories that minimize spatiotemporal variation in areal growth rates and deformation anisotropy. 
The result is a complete prediction for the growth of the surface, including not only a set of intermediate shapes, but also a prediction for cell displacement along those surfaces in the process of growth.
Optimization of growth trajectories for both idealized surfaces and those observed in nature show that relative growth rates can be uniformized at the cost of
introducing anisotropy.
Minimizing the variation of programmed growth rates can therefore be viewed as a generic mechanism for growth pattern selection and may help to understand the prevalence of anisotropy in developmental programs.
\end{abstract}

\maketitle


\section{Introduction}

\begin{figure*}[htbp!] 
  \centering
	\includegraphics{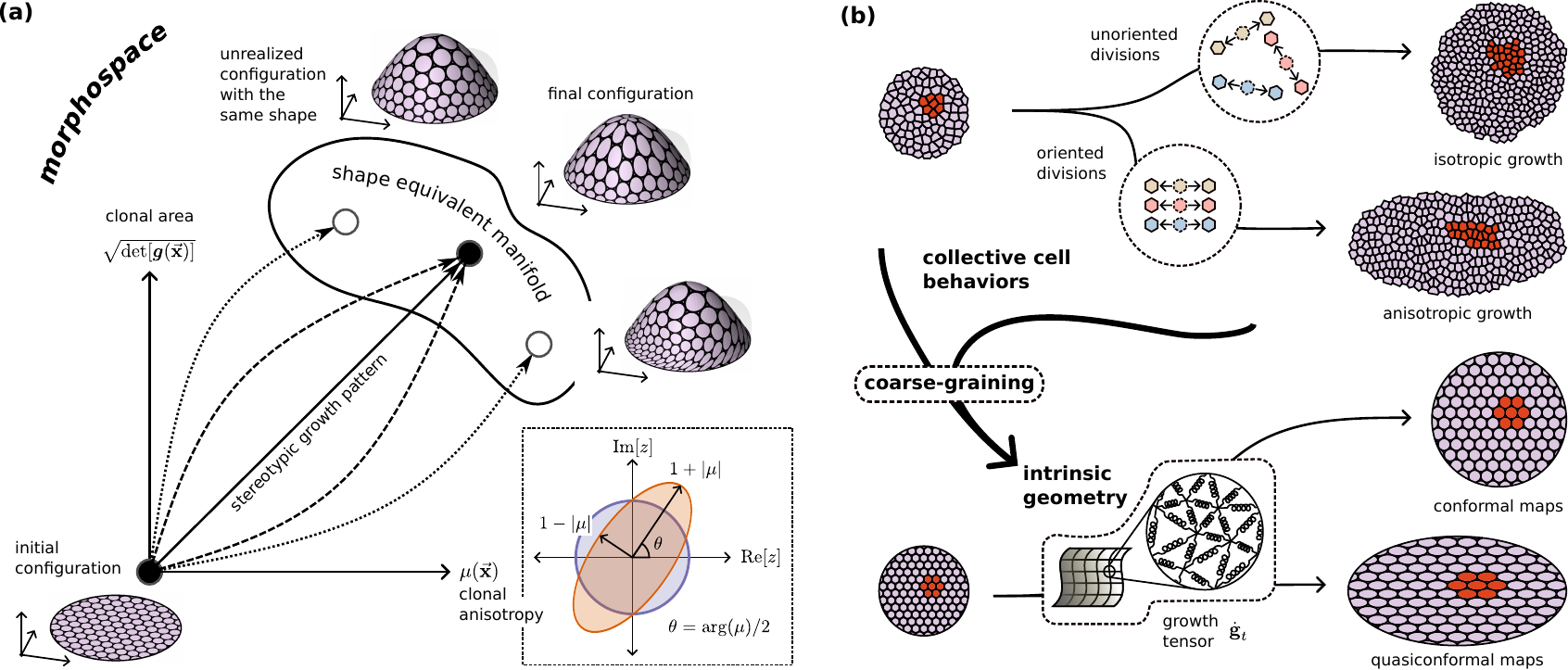}
    \caption{
		(a) Dynamic growth patterns are trajectories in `morphospace'. An arbitrary growth pattern can be reconstructed from the time course of local area and anisotropy in material regions. Different growth patterns may generate the same final shape, but with dissimilar distributions of material patches. Even growth patterns that generate the same final configuration may differ drastically in terms of the realized intermediate configurations. \textit{Inset:} A schematic depiction of how the Beltrami coefficient, $\mu$, encodes the local anisotropy induced by a quasiconformal map.
		(b) Coarse-grained effective theory integrates cellular behaviors into tissue scale deformations. Processes at the cell scale are combined to collectively induce tissue scale shape change. The anisotropy, or lack thereof, of tissue scale growth is determined by the average preferred direction of the underlying cell-scale behaviors. Coarse-graining the cellular motifs yields the \textit{growth tensor}, $\dot{\bf g}_t$, a tensorial representation of the rate and orientation of the time rate of change of the tissue's intrinsic geometry.
		In the schematic, circular patches represent material parcels of tissue that deform along Lagrangian pathlines.
            Dark colored patches highlight growth and deformation of a clonal region.
    }
  \label{fig:growth_and_morphospace}
\end{figure*}

Morphogenesis, the process through which genes generate form, transforms simple initial configurations of cells into complex and specific shapes \cite{Thompson1917,Wolpert2011}.
In order to accomplish this task of self-organization, morphogenetic programs employ both short and long-range molecular signaling, as well as intercellular mechanical interactions, to spatiotemporally coordinate gene expression with cell growth and proliferation \cite{barresi_developmental_2020,Gross2017}.
These interdependent processes all feedback and couple to each other in nontrivial ways.
Controlled cell proliferation and rearrangement forces tissues to deform mechanically, generating specific target architectures selected by evolution from within the enormous `morphospace' of possible shapes and configurations that living systems can assume (\Fig{fig:growth_and_morphospace}(a)).
To help elucidate how developmental processes define the shape of a tissue, we examine the common case of 3D shapes being formed by the growth of 2D tissues, which occurs, for instance, in arthropod limb morphogenesis \cite{Wolff2018,Diaz-de-la-Loza2018} and in plant shoot apical meristem \cite{Hamant2008}.
From a purely geometric perspective, the same target 3D shape can be generated by infinitely many different growth patterns \cite{irvine2017planargrowth,yavari2010growthmech}.
In particular, the ``morphogenetic trajectories'' (i.e. the sequence of growth and local shape changes) of clonal regions defining each growth pattern may be markedly different despite converging to equivalent tissue scale shapes as a whole.
Nevertheless, actual developmental processes are stereotypic across length and time scales, from the behavior of small domains of dividing cells \cite{Foe1989}, to coarse-grained tissue scale flows \cite{Mitchell2022Atlas}, to entire organismal stages of development \cite{bate2009development,Kimmel1995,Browne2005}.
Here we formulate the question of ``growth pattern selection'' by defining possible criteria that distinguish between different spatiotemporal growth programs leading to the same final 3D shape.

We address the problem of shape formation in the context of epithelial morphogenesis.
Epithelia are the fundamental tissue scale building block of multicellular systems and regularly undergo dramatic shape changes during development \cite{Gilmour2017,guillot2013epithelia}.
These tissue scale transformations are driven by a limited repertory of cell behaviors including, most notably, cell growth and division, as well as cell rearrangement and shape change (\Fig{fig:growth_and_morphospace}(b)) \cite{Blanchard2009,Etournay2015,Guirao2015}.
It is intuitively evident that cell division and growth within a 2D tissue layer will tend to increase the layer's area in order to avoid strong lateral compression of cells and concomitant thickening of the epithelial monolayer \cite{Shraiman2005}.
This increase in the tissue area can be accommodated by out-of-plane bending that relieves the resultant in-plane mechanical stress.
3D shape formation due to nonuniform in-plane expansion has been documented in biological morphogenesis (e.g.  in leaf growth \cite{nath2003leaf}) and has also been realized in engineered materials \cite{klein2007shapenem}.
Here, we focus on the shape-forming capabilities of spatially modulated in-plane growth in thin tissues, 
deferring discussion of additional mechanisms of shape control available in thicker tissues or bilayers \cite{VanRees2017}.
Crucially, cell division and tissue growth are often anisotropic.
In addition to specifying a non-uniform profile of areal growth, morphogenetic programs can also prescribe a variable degree of growth anisotropy with preferred orientations that vary throughout the tissue. 
Such anisotropy is a ubiquitous hallmark in the growth and development of plants and animals.  Numerous morphogenetic processes, including organogenesis \cite{BenAmar2013AnistropicGut,Boselli2017,Mitchell2022Gut}, appendage formation \cite{Dye2017,Wolff2018} and vertebrate gastrulation \cite{Gillies2011,Saadaoui2020}, have been shown to feature anisotropic components.

Developmental systems differ not only in their spatiotemporal patterns of growth, but also in the molecular-genetic and cell-biological mechanisms that define and reproducibly establish these patterns.
Our analysis focuses on understanding the general relationship between spatiotemporal patterns of anisotropic 2D growth and 3D shape.
Accordingly, we subsume the details of the underlying cellular processes into a coarse-grained, effective theory that integrates collective cell behaviors into smooth tissue scale deformations.
In this framework, active internal processes within the tissue induce changes to the coarse-grained \textit{intrinsic geometry} (see \Fig{fig:growth_and_morphospace}(b))\cite{yavari2010growthmech,Al-Izzi2021}.
As the intrinsic geometry is updated, the physical geometry morphs via mechanical relaxation in an attempt to match the intrinsically specified target configuration.
This intuitive description is captured mathematically by the formalism of metric elasticity \cite{sharon2010mechanics,efrati2013} extended to morphogenetic dynamics by relating the time derivative of the metric - the mathematical descriptor of \textit{intrinsic geometry} - to anisotropic growth.
Different growth scenarios are distinct in their distributions of clonal regions, each clone being derived by cell division from a single ancestral cell at an early stage of the process.
\Fig{fig:growth_and_morphospace}(b) illustrates the different final configurations produced by an isotropic and an anisotropic growth pattern with the same initial configuration and uniform areal growth rates.
If we allow growth rates to vary in both space and time, we can produce the same final coarse grained shape through either isotropic or anisotropic growth patterns (\Fig{fig:growth_and_morphospace}(a)).
Hence different growth trajectories leading to the same shape may require very different morphogenetic programs to be encoded and executed by the cells. 

We can quite generally link local growth patterns on the scale of cells with the global shape of tissue, by considering incremental maps of the surface, defined explicitly by tracking expanding clone territories over short times. On the coarse grained level, such maps can be generated by (locally expanding) continuous flows. The Lagrangian  morphogenetic trajectories of these flows characterize the behaviors of individual cells and their offspring, thereby distinguishing between the different growth scenarios that lead to identical final shapes. 
This correspondence enables trajectories of physical configurations in morphospace to be quantified in terms of the time dependence of the intrinsic geometry. Flows of cells become smooth time dependent maps. In particular, flows due to isotropic growth are angle-preserving conformal maps, whereas flows due to anisotropic growth are described by quasiconformal maps, i.e. smooth transformations of bounded anistropic distortion \cite{ahlfors2006lectures,lehto1973quasiconformal,gardiner2000quasiconformal}.

Quasiconformal (QC) tranformations provide a natural language for describing anisotropic growth and linking it to surface geometry. Using the mathematical formalism of quasiconformal maps, we demonstrate how complex, nonlinear deformations can be decomposed into sequences of simple infinitesimal updates and thereby linked to infinitesimal  changes to the intrinsic geometry of a thin tissue  expressed in terms of an anisotropic local growth rate.   This will, in turn, allow us to formulate a simple action-like principle for growth pattern selection, given fixed initial and final shapes of the surface. We propose that developmental growth patterns are the `simplest' growth strategies in the sense of requiring the least amount of {\it a priori} information for their specification. Specifically, we consider minimization of the spatiotemporal variation in growth rates and anisotropy.  Optimizing the corresponding functional for given initial and final geometries produces a complete, testable prediction for the full time course of a dynamic growth pattern. Focusing on the case of constant growth, in which growth rates can vary in space, but are held constant in time, we use this ``morphogenetic action'' principle to deduce the optimal growth trajectories for a variety of synthetic shapes. Finally, we apply our approach to deduce the optimal growth trajectories for limb morphogenesis in the crustacean {\it Parhyale hawaiensis}, where developing appendages can be analyzed \emph{in toto} with single-cell resolution \cite{Wolff2015}.

\section{Quasiconformal parameterizations of anisotropic growth patterns}

Let us represent a growing thin tissue with a continuous curved surface $\mathcal{M}_t \subset \mathbb{R}^3$ and assign to each material parcel in the tissue a set of curvilinear Lagrangian coordinates $\vec{\bf x} = (x^1, x^2) \in \mathcal{B}$ defined over a planar domain of parameterization $\mathcal{B} \subset \mathbb{R}^2$. The embedding of the surface in 3D is a time dependent map $\vec{\bf R}(\vec{\bf x}, t):\mathcal{B} \rightarrow \mathbb{R}^3$.  Each point $\vec{\bf R}(\vec{\bf x}, t) \equiv \vec{\bf R}_t(\vec{\bf x}) \in \mathcal{M}_t$ is characterized by its tangent vectors, $\vec{\bf e}_1 = \partial \vec{\bf R} / \partial x^1$ and $\vec{\bf e}_2 = \partial \vec{\bf R} / \partial x^2$, and its unit normal vector $\hat{\bf n} = \vec{\bf e}_1 \times \vec{\bf e}_2 / || \vec{\bf e}_1 \times \vec{\bf e}_2 ||$.
The covariant geometry of the surface is captured by the induced metric tensor
\begin{equation}
	g_{\alpha \beta}(\vec{\bf x}, t) = \frac{\partial \vec{\bf R}_t(\vec{\bf x})}{\partial x^{\alpha}} \cdot \frac{\partial \vec{\bf R}_t(\vec{\bf x})}{\partial x^{\beta}},
	\label{OGMetric1}
\end{equation}
which quantifies lengths and angles between nearby points on the surface.
From this `top-down' perspective, ${\bf g}(\vec{\bf x},t) \equiv {\bf g}_t(\vec{\bf x})$ is a characterization of the surface embedding defined with respect to a given $\vec{\bf R}_t(\vec{\bf x})$.
We can alternatively assume a complementary `bottom-up' perspective in which ${\bf g}_t(\vec{\bf x})$ is instead viewed as a direct description of the underlying local geometry, i.e. the position, size, shape, and arrangement of material regions labelled by the time independent Lagrangian coordinates $\vec{\bf x}$.
Accordingly, we can calculate the 3D embedding $\vec{\bf R}_t(\vec{\bf x})$ in terms of ${\bf g}_t(\vec{\bf x})$, provided some physically plausible assumptions.
Assuming the system behaves as a thin elastic sheet, $\vec{\bf R}_t(\vec{\bf x})$ is determined by minimizing an in-plane stretching energy that tries to match the lengths and angles between material points on the 3D surface with those prescribed by ${\bf g}_t(\vec{\bf x})$ \cite{sharon2010mechanics,efrati2013}.
If the time scale of elastic relaxation in the tissue is short compared to the time scale over which ${\bf g}_t(\vec{\bf x})$ changes, the system will effectively always be in mechanical equilibrium.
For our purposes, we assume that the physical geometry is always an isometric embedding of the target geometry, i.e. the tissue adopts a stress-free equilibrium configuration specified by ${\bf g}_t(\vec{\bf x})$.
Notice that such growth patterns will always be incompressible: new cells introduced into the tissue by cell proliferation increase the local area of tissue patches so that cell density remains constant along Lagrangian pathlines. 
The metric ${\bf g}_t(\vec{\bf x})$ can therefore be used as a representation of the time course of both shape change and material flow along the dynamic surface.
Strictly speaking, even for thin sheets, where the elastic energy is dominated by stretching and compression, additional out-of-plane bending considerations can become relevant to distinguish isometric deformations that do not change lengths and angles in the surface.
A more detailed discussion of the consequences of these effects can be found in \Appendix{app:num:elastic}.


We next relate the time dependence of ${\bf g}_t(\vec{\bf x})$ to growth.
We start by defining a parameterization for the initial surface $\mathcal{M}_0$.
Cumulative changes to the system geometry will be measured relative to this initial configuration.
For simplicity, we restrict our consideration to surfaces with disk-like topology (although virtually identical arguments hold true for both topological cylinders and topological spheres \cite{gardiner2000quasiconformal,lui2012bhf}). 
The Riemann mapping theorem assures us that it is always possible to find a conformal mapping from the unit disk $\mathbb{D} = \left\{ \vec{\bf x} \, : \,||\vec{\bf x}|| < 1\right\}$ to the disk-like surface in 3D, i.e.
$\vec{\bf R}_0(\vec{\bf x}) : \mathbb{D} \rightarrow \mathcal{M}_0$ \cite{needham1998visual,lehto2011univalent}.
This mapping is unique if we constrain the motion of one material point in the bulk of the tissue and another material point on the tissue boundary.
Without loss of generality, we identify these material parcels with the points $\vec{\bf x} = (0,0)$ and $\vec{\bf x} = (1,0)$, respectively.
Identifying our domain of parameterization $\mathcal{B}$ with the unit disk, this mapping endows the surface with a set of so-called \emph{isothermal} coordinates, in terms of which the metric tensor takes the following diagonal form
\begin{equation}
	{\bf g}_0(\vec{\bf x}) = e^{\Omega_0(\vec{\bf x})} \, \boldsymbol{1},
	\label{og:VIP_Metric1}
\end{equation}
where $\boldsymbol{1}$ denotes the rank-2 identity tensor.
Because the mapping is conformal, it preserves angles.
The images of coordinate curves that intersect orthongonally in $\mathcal{B}$ will also intersect orthogonally on the the embedded surface.
Geometrically, we can interpret the `conformal factor', $e^{\Omega_0(\vec{\bf x})}$ as a non-uniform dilation, describing the spatially modulated isotropic swelling that transforms material patches in the domain of parameterization into their corresponding configurations on the physical 3D surface. 

Throughout development, internal biological processes, such as cell growth and proliferation, induce changes to the system's intrinsic geometry. These changes are captured by the time derivative of the metric $\dot{\bf g}$, which we call the \emph{growth tensor}. In general, the growth tensor at some arbitrary time $t$ during the developmental process can be written as $\dot{\bf g}_t(\vec{\bf x}) = \Gamma(\vec{\bf x},t) \, {\bf g}_t(\vec{\bf x}) + \boldsymbol{\gamma}(\vec{\bf x},t)$, where $\text{Tr}[{\bf g}^{-1}_t \boldsymbol{\gamma}] = 0$. The quantity
\begin{equation}
	\Gamma(\vec{\bf x},t) = \frac{1}{2} \, \text{Tr}\left[{\bf g}^{-1}_t \dot{\bf g}_t(\vec{\bf x}) \right] = \frac{d}{dt} \text{log}\left[\sqrt{g_t}\right]
	\label{og:bigGamma1}
\end{equation}
is the relative rate of area change of a local material patch on the 3D surface.
Here we have used the shorthand $g_t = \text{det}[{\bf g}_t]$. The traceless component $\boldsymbol{\gamma}$ is related to anisotropic growth, in a manner that we will shortly make explicit.
The growth tensor can be thought of as introducing infinitesimal updates to the system's instantaneous intrinsic geometry, i.e. ${\bf g}_{t+\delta t}(\vec{\bf x}) = {\bf g}_t(\vec{\bf x}) + \delta t \, \dot{\bf g}_t(\vec{\bf x}) + \mathcal{O}(\delta t^2)$. Each infinitesimal change in the intrinsic geometry will induce a corresponding evolution of the 3D surface. Over time, many infinitesimal updates may combine to produce dramatic 3D shape changes. Consider the growing surface at some later time $\mathcal{M}_t$. For arbitrary growth patterns, the Lagrangian coordinates $\vec{\bf x}$ will no longer necessarily constitute a conformal parameterization at time $t$, despite being initially conformal by construction. However, just as we did with the initial surface, we can construct an instantaneously conformal parameterization of this new surface $\vec{\bf R}_t(\vec{\bf u}) : \mathbb{D} \rightarrow \mathcal{M}_t$, in terms of a new set of \emph{time dependent} virtual isothermal coordinates $\vec{\bf u}$. For consistency, we demand that the material points at $\vec{\bf u} = (0,0)$ and $\vec{\bf u} = (1,0)$ correspond to the same points in the Lagrangian parameterization. In practice, the motion of these points in 3D may be determined by tracking individual cells or other distinguishable features. In general, however, all other material points will be assigned new labels in the virtual parameterization. The key insight is that the virtual isothermal coordinates $\vec{\bf u}(\vec{\bf x},t)$ can be thought of as a transformation of $\mathbb{D}$ into itself, but mapping material points labelled by $\vec{\bf x}$ into new locations. Explicitly, the complete material mapping, describing the true motion of material points in 3D, is given by the composition $\vec{\bf R}_t \circ \vec{\bf u} \circ \vec{\bf R}^{-1}_0: \mathcal{M}_0 \rightarrow \mathcal{M}_t$ where $\vec{\bf u}:\mathbb{D} \rightarrow \mathbb{D}$ is some sense-preserving diffeomorphic reparameterization of $\mathbb{D}$ fixing the points $\vec{\bf x} = (0,0)$ and $\vec{\bf x} = (1,0)$. 

All admissible reparameterizations $\vec{\bf u}(\vec{\bf x}, t)$ are examples of quasiconformal maps (see \Appendix{app:qc}). Quasiconformal mappings are generalizations of conformal mappings that allow for shear deformations of bounded distortion \cite{ahlfors2006lectures}. Their study was effectively inaugurated by Gauss, who used them to investigate the existence of isothermal coordinates on surface patches embedded in 3D \cite{lehto2011univalent}. Let us define a pair of complex variables $z = x^1 + i \,x^2$ and $w = u^1 + i \,u^2$. The map $w(z):\mathbb{D} \rightarrow \mathbb{D}$ is considered quasiconformal if it is a solution to the \textit{complex Beltrami equation}
\begin{equation}
	\frac{\partial w}{\partial \bar{z}} = \mu(z, \bar{z}) \, \frac{\partial w}{\partial z}
	\label{og:beltrami1}
\end{equation}
where where $\bar{z} = x^1-i \,x^2$ and the \textit{Beltrami coefficient}, $\mu$, is subject to $|\mu|_{\infty} < 1$. This last condition is necessary and sufficient to ensure that the Jacobian determinant of the transformation $J = |\partial_z w|^2 - |\partial_{\overline{z}} w|^2 > 0$, such that $w(z)$ is a sense-preserving diffeomorphism. The geometric properties of quasiconformal transformations can be elucidated by comparison with conformal mappings.  Locally, around a point, a conformal transformation maps infinitesimal circles into similar circles.  Therefore, a conformal map does not introduce any preferred local orientation and is necessarily isotropic.  In contrast, a quasiconformal transformation maps infinitesimal circles into similar ellipses.  The Beltrami coefficient encodes both the magnitude and directionality of this distortion and, thus, also the anisotropy of the transformation.  This geometric intuition is illustrated in the inset of \Fig{fig:growth_and_morphospace}(a). Notice that when $\mu = 0$, the Beltrami equation reduces to the Cauchy-Riemann equation, $\partial_{\overline{z}} w = 0$, the necessary and sufficient condition for the conformality of a complex map. The Measurable Riemann Mapping Theorem assures us that there is a unique solution $w(z)$ to the complex Beltrami equation (\Eq{og:beltrami1}), fixing the points $z = 0$ and $z = 1$, for any $\mu$ with $|\mu|_{\infty} < 1$ \cite{Astala2009}. More information about quasiconformal mappings can be found in \Appendix{app:qc}.

\begin{figure*}[htbp!]
	\centering
	\includegraphics{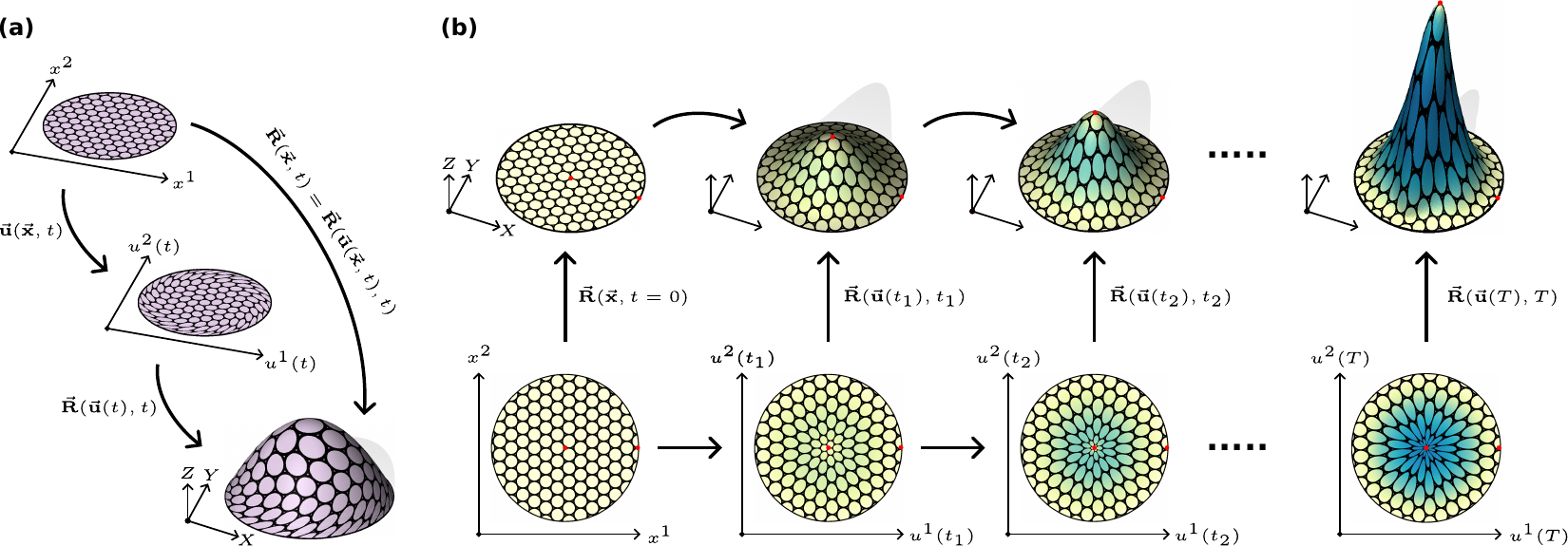}
	\caption{
	Quasiconformal parameterization of dynamic growth patterns.
	(a) An arbitrary parameterization of a topological disk in 3D can be expressed as the composition of a quasiconformal automorphism of the unit disk $\vec{\bf u}:\mathbb{D} \rightarrow \mathbb{D}$, fixing the points $\vec{\bf x} = (0,0)$ and $\vec{\bf x} = (1,0)$, followed by a conformal mapping into 3D, $\vec{\bf R}:\mathbb{D}\rightarrow \mathbb{R}^3$. The parameterization is unique if both a Beltrami coefficient $\mu$ and the motion of two material points are specified.
	(b) Complicated time-dependent growth patterns for 3D surfaces can be built by stringing together simple infinitesimal updates to the system's intrinsic geometry. Beginning with a conformal parameteriztion of the initial configuration, new configurations are generated by calculating the contributions of changes in the system's intrinsic anistropy and target areas. A time course of $\dot{\mu}$ and $\Gamma$ are sufficient to fully reconstruct arbitrary growth patterns.
	}
	\label{fig:og:qc_dynamic_param}
\end{figure*}

Any arbitrary time dependent arrangement of material points on a growing 3D surface can therefore be captured by the composition  $\vec{\bf R}_t \circ w(z,t) : \mathbb{D} \rightarrow \mathcal{M}_t$.
All of the anisotropy in the parameterization is contained within the intermediate quasiconformal transformation $w$ and is described by the associated Beltrami coefficient $\mu = \partial_{\bar{z}} w / \partial_z w$.
This construction is illustrated in \Fig{fig:og:qc_dynamic_param}(a).
In terms of these quantities, the complete Lagrangian metric tensor is given by
\begin{equation}
	{\bf g}_t(z) = e^{\Omega \circ w} |\partial_z w|^2 \,
	\begin{pmatrix}
		|1+\mu|^2 & -i(\mu - \bar{\mu}) \\
		-i(\mu - \bar{\mu}) & |1-\mu|^2
	\end{pmatrix},
	\label{og:QC_Metric1}
\end{equation}
expressed in the real basis of the Lagrangian coordinates $\vec{\bf x}$. Notice that, given a 3D surface and the motion of a single point on the boundary and a single point in the bulk, each $\mu$ corresponds to a unique parameterization of that surface thanks to the uniqueness properties of both $\vec{\bf R}_t$ and $w(t)$ (see \Appendix{app:qc}).

The next crucial step is to explicitly determine how growth induces changes to this quasiconformal geometry. 
At an arbitrary time $t$ during the developmental process, the metric in \Eq{og:QC_Metric1} expressed in the virtual isothermal coordinates has the simple diagonal form ${\bf g}_t(w) = e^{\Omega(w,t) } \, \boldsymbol{1}$.
The most general possible growth tensor in these coordinates has the following form
\begin{equation}
    \begin{split}
        &\dot{\bf g}_t(w) =  \Gamma(w,t) \, {\bf g}_t(w) + \boldsymbol{\gamma}(w,t) \\
        &= \Gamma(w,t) \, e^{\Omega(w,t)} \,  \boldsymbol{1} \\
        &+ 2\, e^{\Omega(w,t)} \, |\gamma(w,t)| \,
        \begin{pmatrix}
            \cos{ \vartheta(w,t)} &  \sin{ \vartheta(w,t)} \\
            \sin{ \vartheta(w,t)} & -\cos{ \vartheta(w,t)} 
        \end{pmatrix}
        ,
    \end{split}
    \label{og:dot_Metric}
\end{equation}
expressed in the real basis of the intermediate coordinates $\vec{\bf u}$. Here, $\gamma(w,t) = |\gamma(w,t)| \, e^{ i \vartheta(w,t)}$ is a complex field describing the magnitude and orientation of infinitesimal anisotropic updates to the intrinsic geometry. Clearly, for $\gamma = 0$, $\Gamma$ simply modulates the conformal factor, whereas non-zero $\gamma$ induces local shearing. The infinitesimal update to the geometry shown in \Eq{og:dot_Metric} will induce a corresponding evolution of the intermediate isothermal coordinates
\begin{equation}
    w(z,t+\delta t) = w(z,t) +  h_t \circ w(z,t) \, \delta t + \mathcal{O}(\delta t^2),
    \label{og:inf_VIP1}
\end{equation}
The infinitesimal mapping $h_t$ can be thought of as the instantaneous `velocity' of material points in the isothermal coordinates. Simultaneously, there will also be a change in the overall 3D shape of the surface captured by the conformal factor. The total relative rate of area change on the 3D surface resulting from these updates can be expressed in the intermediate coordinates as
\begin{equation}
    \begin{split}
        &\Gamma(w,t) = \frac{d \,{\Omega}(w,t)}{dt} + 2 \, \text{Re}\left[\partial_w h_t(w,t)  \right] \\
        &= \partial_t \Omega(w,t) + 2 \, \text{Re}\left[ h_t(w,t) \, \partial_w \Omega(w,t) \right] + 2 \, \text{Re}\left[\partial_w h_t(w,t) \right].
     \end{split}
    \label{og:Omega_Dot_VIP}
\end{equation}
We see that $\Gamma$ is defined by contributions from both the change of the overall surface shape and the redistribution of points on the surface. The infinitesimal anisotropy of the mapping is simply $\partial_{\bar{w}} w(t+\delta t)/\partial_w w(t+\delta t) \approx \partial_{\bar{w}} h_t = \gamma(w,t)$.

In order to complete the description, we must relate the infinitesimal geometric updates in the intermediate frame to their counterparts in the Lagrangian coordinates.
Application of the chain rule tells us that the the time rate of change of the cumulative anisotropy is given by
\begin{equation}  
    \dot{\mu}(z, t) = (1-|\mu(z,t)|^2) (\gamma \circ w (z,t)) \,e^{i\psi},
    \label{bhf_mu1}
\end{equation}
where $e^{i\psi} = \overline{\partial_z w}/\partial_z w$. According to \Eq{bhf_mu1}, the same $\gamma$ will produce diminished transformations of the cumulative geometry in regions that are already highly anisotropic.
When expressed directly in Lagrangian coordinates, the infinitesimal mapping $h_t[w, \dot{\mu}](z,t)$ becomes a functional of the intermediate parameterization $w$ and  $\dot{\mu}$. The explicit construction for  $h_t[w, \dot{\mu}]$ is called the \emph{Beltrami holomorphic flow} (BHF) \cite{gardiner2000quasiconformal} and is discussed in detail in \Appendix{app:qc:BHF}.
Finally, direct calculation tells us that
\begin{equation}
    \begin{split}
        &(\Gamma \circ w)(z,t) = \frac{d}{dt} (\Omega \circ w) \\
        &\qquad + 2 \, \text{Re}\left[ \frac{(\partial_z h_t[w, \dot{\mu}]) \overline{\partial_z w}}{|\partial_z w|^2} \right] + 2 \, \text{Re}\left[\frac{\dot{\mu} \, \bar{\mu}}{1-|\mu|^2} \right],
     \end{split}
    \label{og:Omega_Dot_Lagr}
\end{equation}
in which form it is explicitly clear that local area change on the 3D surface in Lagrangian material patches depends on cumulative anisotropy.

We are now equipped with all of the necessary machinery to quantify dynamic growth patterns in thin tissues. At time $t = 0$, we can construct a conformal Lagrangian parameterization of the initial surface. A short time later, internal growth processes update the intrinsic anisotropy of the system according to $\dot{\mu}$. This change induces a corresponding flow of the quasiconformal mapping captured by the quantity $h_t[w, \dot{\mu}]$. Simultaneously, these processes also induce a change in the 3D area of material patches, which is captured by the quantity $\Gamma$. We can calculate a corresponding update to ${\bf g}_t$ in terms of $\dot{\mu}$ and $\Gamma$. This geometry can then be embedded into 3D to find the new configuration of the system. We can string these infinitesimal transformations together sequentially to construct any arbitrary growth trajectory. This construction is illustrated in \Fig{fig:og:qc_dynamic_param}. Therefore, if we know the time course of cumulative anisotropy $\mu$ and the 3D area $A$ of each material patch, we can uniquely construct the associated metric tensor, which serves as a proxy for both shape and flow.  This decomposition enables us to assign a simple, concrete structure to the abstract morphospace alluded to into in the introduction. Morphogenetic trajectories of thin tissues are simply time dependent profiles of anisotropy and areal growth (\Fig{fig:growth_and_morphospace}(a)).

\section{Growth pattern selection as an optimization problem}

As we have already stated, there is an infinite degeneracy of possible growth trajectories linking an initial configuration and a final shape.  How does nature settle on a specific developmental program? To explore the possibility that natural selection acts as constrained optimization, we can frame the growth pattern selection as a variational problem. Given an initial configuration $\mathcal{M}_0$ (interpreted as a fixed initial shape \emph{and} a fixed initial parameterization) and a final shape $\mathcal{M}_T$ (a fixed final shape, but \emph{not} a fixed final parameterization), we define the optimal growth trajectory $\dot{\bf g}^*_t(\vec{\bf x}) = \argmin_{\dot{\bf g}_t(\vec{\bf x})} \mathcal{S}$ as the minimizer of
\begin{equation} 
    \begin{split}
	    \mathcal{S} &= \int_0^T dt \, \int_{\mathcal{B}} d^2\vec{\bf x}\, \sqrt{g_t} \, \mathcal{L}\left[ \dot{\bf g}_t(\vec{\bf x}); \mathcal{S}(t=0) \right]\\
            &\qquad \qquad \qquad \qquad + \lambda \, D_{SEM}[{\bf g}_T; \, \mathcal{M}_T].
	\end{split}
	\label{og:OptimalGrowth2}
\end{equation}
In the language of optimal control theory \cite{Betts2010OC}, the first term is the `cost functional' of the growth pattern calculated in terms of a running cost  density $\mathcal{L}$. The second term is a Lagrange multiplier enforcing the constraint that the metric tensor at the final time ${\bf g}_T(\vec{\bf x})$ must be a valid induced metric of the final surface $\mathcal{M}_T$. There is also an additional implicit constraint that ${\bf g}_t$ remain a valid metric for all $t$, i.e. ${\bf g}_T$ is a symmetric, positive-definite type-(0,2) tensor.

This shape constraint enforced by the Lagrange multiplier demands some additional explanation. We say that two metrics ${\bf a}$ and ${\bf g}$ are shape equivalent metrics (SEMs) if they are both induced metrics of the \emph{same} surface $\mathcal{M}_T$, but corresponding to \emph{different} coordinate parameterizations. For notational convenience, we denote by $\Sigma_{\mathcal{M}_T}$ the space of all SEMs for the surface $\mathcal{M}_T$. Physically, given a shared Lagrangian parameterization $\vec{\bf x}$ defined using $\mathcal{M}_0$, we interpret these parameterizations as resulting from different material flows encapsulated by contrasting intermediate mappings $\vec{\bf u}(\vec{\bf x}, T)$. If we were to fix the complete configuration of the final surface, rather than just its shape, we would over-constrain the entire growth trajectory since intermediate configurations are linked to the final configuration via the machinery discussed in the previous section. For instance, we would not be able to objectively investigate the conditions under which growth pattern selection criteria give rise to anisotropy, since anisotropy (or lack thereof) would be baked into the final configuration by construction. Therefore, in order to generate an unbiased answer to the question of optimal growth, it is critical that we allow ${\bf g}_T$ to search the entire space of $\Sigma_{\mathcal{M}_T}$.  Anything less results in overly restrictive demands on the material flows during growth.

Determining whether or not ${\bf g}_T \in \Sigma_{\mathcal{M}_T}$ therefore requires searching the space of possible parameterizations of $\mathcal{M}_T$. In general, the complicated nature of this functional space makes problems of this type quite challenging. Recall, however, that each diffeomorphic parameterization of a disk-like surface corresponds to a unique Beltrami coefficient, so long as we constrain the motion of two material points. We can therefore simply explore the space of Beltrami coefficients, rather than directly searching the space of surface diffeomorphisms. This amounts to a much simpler task, since there are no restrictions that Beltrami coefficients must be one-to-one, onto, or satisfy any constraints on their Jacobians \cite{lui2012bhf}. Explicitly, we have
\begin{equation} 
	D_{SEM}[{\bf g}_T; \, \mathcal{M}_T] = \min_{\mu} \, \frac{1}{2} \, \int_{\mathbb{D}} ||{\bf g}_T - {\bf a}[\mu]||^2 \, \sqrt{g_T} \, d^2\vec{\bf x},
	\label{og:DSEM2}
\end{equation} 
where $||\cdot||$ denotes the Frobenius norm and ${\bf a}[\mu] \in \Sigma_{\mathcal{M}_T}$ is the induced metric of $\mathcal{M}_T$ generated by $\mu$. The functional $D_{SEM}[{\bf g}_T; \, \mathcal{M}_T]$ will vanish if and only if ${\bf g}_T \in \Sigma_{\mathcal{M}_T}$. Formulating the nontrivial shape constraint in this way transforms it into a variational sub-problem that can be solved by using the BHF to calculate the variation in $D_{SEM}[{\bf g}_T; \, \mathcal{M}_T]$ under the corresponding variation in $\mu$. In practice, this task can be addressed using standard gradient-based optimization methods \cite{nocedal2006numerical}.

We now return to the specific form of the cost functional defining optimal growth trajectories. We propose that optimal growth patterns minimize spatiotemporal variation in growth rates and anisotropy, i.e.
\begin{equation}
    \begin{split}
	    \mathcal{S} &= \lambda \, D_{SEM}[{\bf g}_T; \, \mathcal{M}_T] \quad + \\
	    &\int_0^T dt \, \int_{\mathcal{B}} d^2\vec{\bf x}\, \sqrt{g_t} \, \, \left[ c_1 |\nabla {\Gamma}|^2 + c_2 |\nabla \gamma|^2 + c_3 \dot{\Gamma}^2 + c_4 |\dot{\gamma}|^2  \right].
	\end{split}
	\label{og:mingro1}
\end{equation}
We use
$\nabla$ to denote the covariant derivative of a quantity with respect to the metric ${\bf g}_t$ in terms of the  fixed 2D Lagrangian coordinates $\vec{\bf x}$.
For simplicity and expediency, we focus on a the special case of temporally \emph{constant growth} corresponding to the limit of high penalties ($c_3,c_4$) for temporal variation of $\Gamma$ and $\gamma$.
Thus $\Gamma$ and $\gamma$ may vary spatially, but are held constant for all time, i.e. $\dot{\Gamma} = \dot{\gamma} = 0$.
Notice constant growth implies that the area of each material patch grows exponentially with a time-independent rate constant. 
We can clearly see this by calculating the time-dependent area of a material element $dA(\vec{\bf x}, t) = \sqrt{g_t} \, d^2\vec{\bf x}$. From \Eq{og:bigGamma1}, we have
\begin{equation}  
	\sqrt{g_t} = e^{\Gamma \, t} \, \sqrt{g_0} \implies dA(t) = e^{\Gamma \, t} dA(t= 0),
	\label{og:cg_area1}
\end{equation}
where we have exploited the fact that $\Gamma$ is constant in time. 
The final simplification is to define the covariant derivatives and area elements with respect to the metric at the initial time ${\bf g}_0$.
This choice completely removes any time dependence from the cost functional.
The modified constant growth functional is given by
\begin{equation}
    \begin{split}
        \mathcal{S} &= \lambda \, D_{SEM}[{\bf g}_T; \, \mathcal{M}_T] \quad + \\
	    &T \int_{\mathcal{B}} d^2\vec{\bf x}\, \sqrt{g_0} \, \, \left[  c_1 |\nabla {\Gamma}|^2 + c_2|\nabla \gamma|^2 \right].
    \end{split}
    \label{og:cg_mingro2}
\end{equation}
Essentially, we assume that the developmental procedure is programmed into the tissue at some initial time and optimization of our cost functional selects the plan that minimizes the complexity needed to specify this primordial growth field.
Solving this constrained problem yields a time course of metric tensors in terms of $\Gamma$ and $\gamma$, which we can them dynamically embed in $\mathbb{R}^3$.
The final result is a complete prediction for both the coarse-grained shape and flow of a growing tissue over time.

\section{Results}

We now apply our formalism to compute the optimal constant growth patterns for a variety of synthetic and natural examples. To that end, we generate numerical solutions using a combination of discrete differential geometry processing and nonlinear optimization methods.  The target geometries produced by our numerical optimization machinery can be embedded sequentially in 3D to reconstruct a full time course of shape and flow in growing surfaces. Details of the optimization methods, the discretization of the BHF, and the methods used to embed optimal intrinsic geometries in 3D can be found in \Appendix{app:num}. In all of the experiments that follow, we set the constants $c_1 = c_2 = 1$.

\begin{figure}[htbp]
    \centering
    \includegraphics{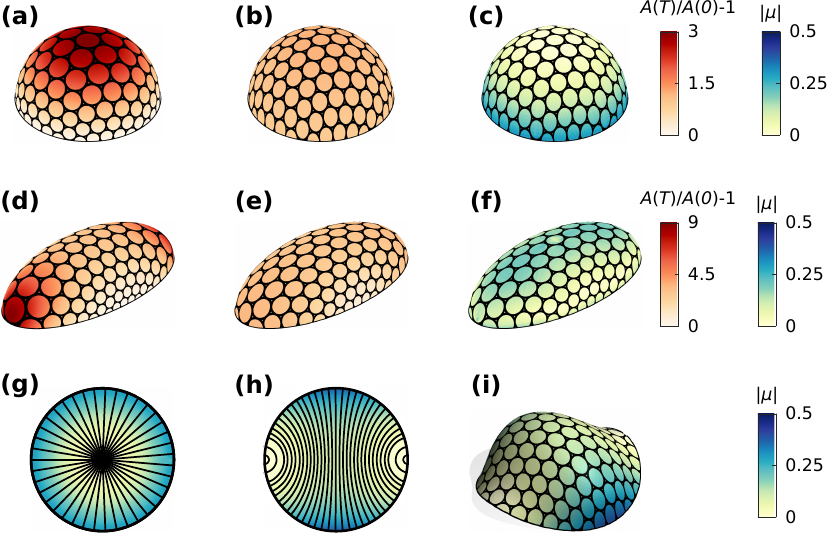}
    \caption{
        \textbf{Growth rates are uniformized by introducing anisotropy.} Optimized constant growth patterns for set of synthetic surfaces. Initial configuration for all surfaces is the flat unit disk. Circular texture on the surfaces represents the deformation of material patches under the flow.
        (a) A conformal growth pattern linking the unit disk and the hemispherical surface. Growth occurs primarily at the apex of the dome.
        (b) The optimized constant growth rates for the same final shape. Growth rates are heavily uniformized relative to the conformal growth pattern.
        (c) The absolute value of the Beltrami coefficient for the optimized constant growth pattern.
        (d) A conformal growth pattern linking the unit disk and an elliptic paraboloid. Growth occurs primarily at the poles.
        (e) The optimized constant growth rates for the same final shape. Growth rates are uniformized, although not as completely as the case of the spherical surface.
        (f) The absolute value of the Beltrami coefficient for the optimized constant growth pattern.
        (g) The absolute value of the Beltrami coefficient for the optimized constant growth pattern transforming a flat disk into a hemispherical surface and the corresponding nematic field illustrating the orientation along which material patches extend. The texture contains a +1 topological defect with phase $\psi = 0$.
        (h) A synthetic Beltrami coefficient with the same total integrated anisotropy constructed by placing two +1 defects with phase $\psi = \pi/2$ at opposite poles of the disk boundary.
        (i) A new surface generated using the exact same growth rates as the surface in (b), but with the anisotropy texture shown in (h).
    }
    \label{fig:og:synth_surf}
\end{figure}

We begin by calculating optimal growth patterns for a set of simple synthetic target shapes grown from an initially flat disk configuration (\Fig{fig:og:synth_surf}). In particular, we solve the optimal growth problem for a hemisphere (\Fig{fig:og:synth_surf}(a-c)) and an elliptic paraboloid (\Fig{fig:og:synth_surf}(d-f)) using a conformal parameterization as our initial guess. For all systems studied, growth rates in the optimal configuration are uniformized at the cost of introducing anisotropy. These results suggest that growth rate uniformization may be a generic mechanism underlying the presence of anisotropy in observed developmental growth patterns.

Our formalism enables us to decode the contributions of areal expansion and anisotropic deformation towards determining 3D shape. \Fig{fig:og:synth_surf}(c) shows the cumulative anisotropy of the optimal constant growth pattern transforming a flat disk into a hemispherical shell. Displaying the local orientation along which tissue parcels extend due to anisotropic deformation results in a nematic texture supporting topological defects (\Fig{fig:og:synth_surf}(g)). Such textures have recently been studied in the context of epithelial morphogenesis coupled to active nematic biomolecular components \cite{Doostmohammadi2018}, where defects in nematic textures have been identified as organizing centers of curvature and mass accretion leading to shape change \cite{Pearce2020,Vafa2022Hydra,Hoffmann2022}. The optimal growth texture is characterized by a $+1$ defect with phase $\psi = 0$ at the pole of the sphere. \Fig{fig:og:synth_surf}(h) displays a modified texture with the same total integrated anisotropy, i.e. $\int_{\mathbb{D}} |\mu(\vec{\bf x})|^2 d^2\vec{\bf x}$, but with an alternative texture constructed by placing two $+1$ defects with phase $\psi = \pi/2$ at opposite poles on the disk boundary. Simulating a growth process using this modified anisotropy texture, but keeping the areal growth rates exactly the same as in \Fig{fig:og:synth_surf}(b), produces an entirely new final shape (\Fig{fig:og:synth_surf}(i)). This proof-of-principle demonstrates how our methods can be used as a platform for growth pattern design in terms of nematic textures.

\begin{figure}[htbp!]
\centering
\includegraphics{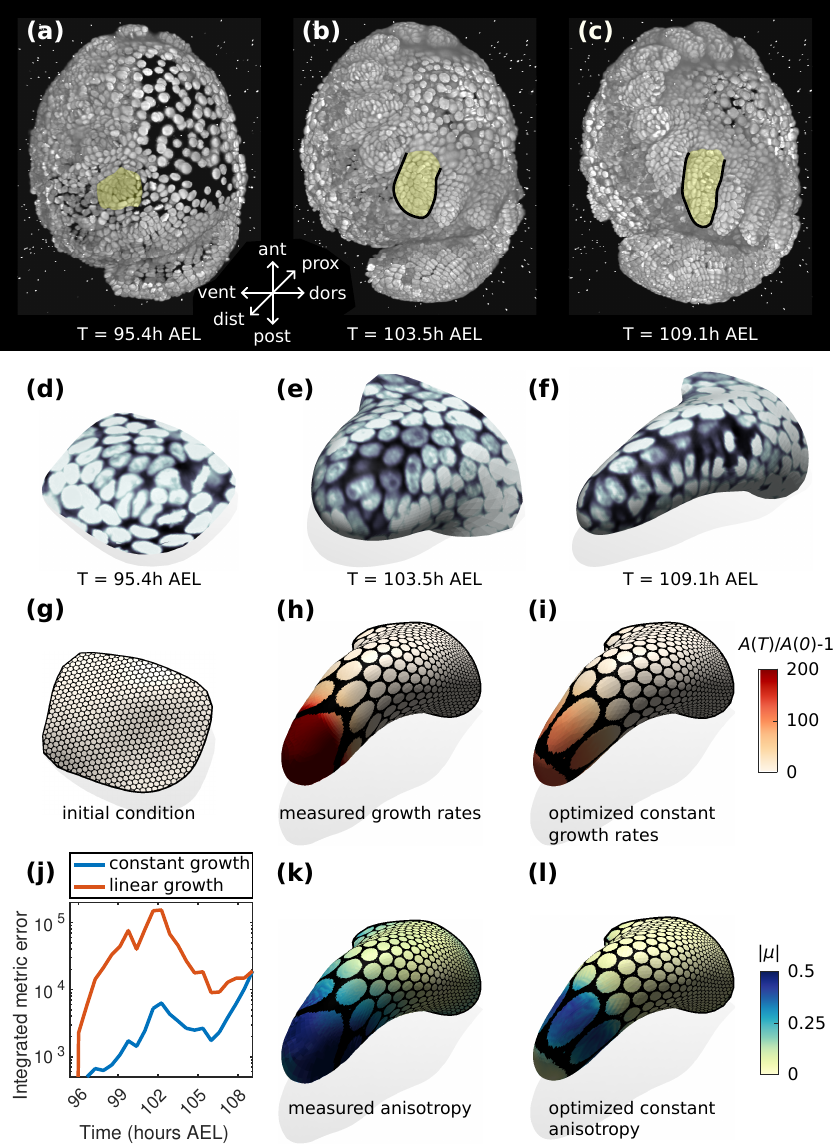}
	\caption{
		\textbf{Optimal growth predictions captures features of experimentally quantified limb morphogenesis.}
		(a)-(c) A growing appendage in a {\it Parhyale hawaiensis} embryo with a fluorescent nuclear marker captured using light-sheet microscopy. Time is measured in hours after egg lay (AEL). Cells corresponding to the growing limb are highlighted in yellow.
		(d)-(f) The surface of the growing limb extracted using tissue cartography.
		(g) Initial condition for the coarse-grained growth pattern. Circular patches are schematic representations of clonal regions.
		(h) The measured areal growth rates during appendage outgrowth.
		(i) The areal growth rates of the optimized constant growth pattern generating the final appendage shape.
		(j) Instantaneous integrated metric error between two different growth patterns. The blue curve is the prediction error for the constant growth pattern linking the initial configuration and the optimized final configuration holding $\dot{\mu}$ and $\Gamma$ constant. The orange curve links the same initial and final configurations, but does so by linearly interpolating the initial and final target geometries, i.e. $\tilde{\bf g}_t = (1-t/T) \, {\bf g}_0 + (t/T) \, {\bf g}_T$ for $t \in [0, T]$.
		(k) The measured anisotropy during appendage outgrowth.
		(l) The anisotropy of the optimized constant growth pattern generating the final appendage shape.
	}
	\label{fig:og:parhyale_leg}
\end{figure}

Next, we applied our machinery directly to experimentally observed pattern of limb growth in the crustacean \textit{Parhyale hawaiensis} (\Fig{fig:og:parhyale_leg}). {\it Parhyale} are direct developers that grow externally visible appendages and deep surface folds during embryogenesis when live imaging is readily possible \cite{Wolff2015,Browne2005}. They are a uniquely suitable system for \emph{in vivo} investigation of the transformation of the 2D limb primordium into its 3D shape, in real time and across multiple length scales \cite{Rallis2022}. Recent work characterized appendage growth by tracking individual cells in 3D \cite{Wolff2018}. Analysis of individual cell behaviors is informative, but, on its own, produces limited insights about the dynamics of tissue-scale deformations. Using a combination of computer vision and discrete differential geometry, we extracted the continuous, quasiconformal growth pattern generating the limb \emph{in vivo}. We were then able to apply our optimization approach to predict growth pattern linking the observed initial configuration and the final shape of the limb. The optimal growth pattern successfully reproduces the qualitative large scale features of biological limb morphogenesis. In both the measured and calculated growth patterns, areal growth occurs primarily at the distal tip of the nascent appendage. The anisotropic deformations that extend the growing limb along its proximodistal axis are also similar. For both the measured and calculated growth patterns, the tissue stretches along the proximodistal axis away from the animal's body.  The most obvious quantitative discrepancy is that the optimal growth pattern under-predicts the maximum areal growth rates at the distal tip of the limb.  Notice that the \textit{Parhyale} limb is comprised of relatively few cells ($\sim 15$ cell lengths long at 109.1h AEL). The observed discrepancy may indicate that the optimal growth principle should be modified to account for the discrete cellular structure of such tissues, rather than naively optimizing for smooth gradients of proliferation. 

Finally, we made quantitative comparisons of different types of dynamic growth patterns to the observed developmental trajectory (\Fig{fig:og:parhyale_leg}(j)).  In particular, we compared the optimal constant growth pattern to a synthetic growth pattern that linearly interpolates between the initial and final optimal metric tensor, i.e. $\tilde{\bf g}_t = (1-t/T) \, {\bf g}_0 + (t/T)\, {\bf g}_T$ for $t \in [0, T]$. We see that the constant growth pattern produces a significantly more accurate results at intermediate times than the linearly interpolated geometry.  

\section{Discussion}

This work establishes a theoretical and computational approach to the study of 3D shape formation by growing 2D (epithelial) sheets. We have framed the problem of morphogenetic growth pattern selection and defined a general action-like variational principle that allows one to determine spatiotemporal growth trajectories that uniformize growth, subject to the constraint of achieving the desired final shape. Our approach enables quantitative comparisons between different growth patterns, opening the door to a predictive understanding of how morphogenetic programs decode genetic information into shape and form.

Our numerical implementation of the variational principle has demonstrated, quite generally, that growth rates may be uniformized at the cost of introducing anisotropy.
Furthermore, already the simplest form of the optimization cost functional was shown to reproduce important qualitative features of experimentally measured growth patterns in
arthropod limb morphogenesis.
These results suggest that growth rate uniformization may serve as a generic mechanism contributing to the prevalence of anisotropy in observed developmental programs. 

Crucially, our optimization principle is both modular and
adaptable. In the simple form used in this work, the optimization cost functional suppresses temporal variation so that the (in general anisotropic) growth rate tensor of a small patch of cells is directly related to the growth rate tensor of its clonal progenitor patch. Optimization in this case is uniformizing the pattern of growth "prescribed" over the initial primordium, thus minimizing the information supplied at the initial time. 
A compelling generalization of this simplest optimization condition would be the inclusion of mechanical and geometric feedback \cite{AlMosleh2018}, which would effectively allow additional factors, e.g. local curvature or stress, to control or modulate morphogenetic growth. In plant shoot apical meristem, for instance, it is known that anisotropic stresses sustained within the outermost epidermal layer adaptively orient stress-bearing microtubules that sculpt the 3D shape of the developing organ by modulating the directions in which cells grow and divide \cite{Hamant2008}.
In the developing \emph{Drosophila} wing, essentially uniform growth rates are maintained through a negative mechanical feeback loop where reduced tension along cell junctions in fast growing clones leads to down regulation of growth \cite{Pan2016}.
By relaxing the constraint that the physical geometry be an isometric embedding of the target geometry, we could directly explore how dynamical mechanical fields, such as in-plane stress, may modulate the feasibility of particular growth patterns with respect to optimality.
Another possibile generalization would be the explicit inclusion of morphogen fields.
Our machinery is well suited to analyze a variety of geometrically distinct classes of morphogens, including scalar fields (e.g. molecular concentration), vector fields (e.g. concentration gradients), and nematic tensor fields (e.g. contractile actomyosin networks).
Essentially arbitrary interactions between mechanics, geometry, and signaling could be incorporated into novel optimality criteria. These explorations would be useful in establishing a link between the pattern formation frequently observed in morphogen signaling and the 3D shapes of organs and appendages.

It would also be interesting to investigate sheets of finite thickness, where additional mechanical control over 3D shape can be provided by direct determination of local bending.
Functionally, this could be executed in our methodology by expanding our simple description of intrinsic geometry to include a time dependent target tissue curvature.
This target curvature may be determined biologically through differential expansion/contraction of the apical and basal surfaces of polarized epithelia, such as the apical constriction of cells that drives ventral furrow formation in \emph{Drosophila} \cite{Noll2017}.
Similar mechanisms are available in the case of a tissue bilayer \cite{VanRees2017}, i.e the differential growth of the adaxial and abaxial sides of leaves \cite{VanDoorn2003}.
Finally, the portability of our methodology invites applications beyond biological morphogenesis to the design of synthetic surface structures and bioinspired materials.

For simplicity, in this work, we did not include an explicit description of individual cells, choosing instead to subsume these behaviors into our continuum scale description of tissues as a whole.
The discrete nature of cells is, however, vital to morphogenesis, insofar as it allows for modes of inter-cell communication and interaction inaccessible to featureless continuous media.
The size of cells also sets a reference scale below which it is impossible or irrelevant to define spatially varying fields of growth parameters.
Many growing tissues are comprised of relatively few cells, but are still amenable to the type of analysis explored in this work (the \emph{Parhyale} limb we analyzed is only $\sim 15$ cell lengths long at 109.1h AEL).
Following recent work \cite{Grossman2022}, the most straightforward way to explicitly include cell behaviors in our formalism would be to retain a continuum description, with material points corresponding to subcellular parcels of tissue, and impose cellular topology on top of this description as a separate, time dependent field.
New optimality criteria could be generated that explicitly depend on cellular topology.
This inclusion would enable the systematic exploration of multi-scale behaviors underlying thin tissue morphogenesis, while retaining the explanatory power of our geometric formalism.

\begin{acknowledgments}
This work was supported by the NSF through PHY 2210612 award. DJC acknowledges support from the NSF through PHY 2013131. We thank Sebastian Streichan for stimulating discussions and critical feedback.
\end{acknowledgments}
\vfill

\appendix
\appsection{A Primer on Quasiconformal Maps}
\label{app:qc}

\subsection{Geometric Properties of Quasiconformal Maps}
\label{app:qc:geo}

This section contains a short mathematical introduction to planar quasiconformal mappings.  Much of this exposition is drawn directly from Ahlfors' extremely lucid monograph \cite{ahlfors2006lectures} and a more comprehensive presentation can be found therein. The structure of a quasiconformal map can be deduced from the familiar notion of a conformal map by the by allowing the transformation to include shear deformations.   Let $w = f(z): \mathbb{C}\rightarrow\mathbb{C}$ be a $C^1$ homeomorphism from one subregion of $\mathbb{C}$ to another given in terms of the complex variables $z = x^1+i x^2$ and $w = u^1 + i u^2$.  The Jacobian determinant of this transformation is
\begin{equation}
	J = |\partial_z w|^2 - |\partial_{\overline{z}} w|^2,
    \label{qc1}
\end{equation}
where we denote the complex conjugate $\overline{z} = x^1 - i x^2$ and the complex derivative operators are defined via the chain rule as
\begin{equation}
	\frac{\partial}{\partial z} = \frac{1}{2} \left( \frac{\partial}{\partial x^1} - i \frac{\partial}{\partial x^2} \right), \qquad \frac{\partial}{\partial \overline{z}} = \frac{1}{2} \left( \frac{\partial}{\partial x^1} + i \frac{\partial}{\partial x^2} \right).
\label{qc2}
\end{equation}
We limit ourselves to the case of sense-preserving diffeomorphisms, for which $J$ is strictly positive, i.e. $|\partial_z w| > |\partial_{\overline{z}} w|$.  Locally, around a point $z_0$, this diffeomorphism induces a linear mapping of the differentials
\begin{equation}
	\begin{split}
    	du^1 &= \partial_{x^1} u^1 \, dx^1 + \partial_{x^2} u^1 \, dx^2 \\
		du^2 &= \partial_{x^1} u^2 \, dx^1 + \partial_{x^2} u^2 \, dx^2,
    \end{split}
    \label{qc3}
\end{equation}
or in terms of the complex notation
\begin{equation}
	dw = \partial_z w \, dz + \partial_{\overline{z}} w \, d\overline{z}.
\label{qc4}
\end{equation}
Geometrically, this is a local affine transformation bringing infinitesimal circles centered about $z_0$ into similar ellipses.  We define the linear distortion of the mapping, $D_w$, as the ratio of the major axis of this image ellipse to its minor axis.  From Equation \eqref{qc4}, it is immediately apparent that
\begin{equation}
	\left( |\partial_z w| - |\partial_{\overline{z}} w| \right) |dz| \leq |dw| \leq \left( |\partial_z w| + |\partial_{\overline{z}} w| \right) |dz|
\label{qc5}
\end{equation}
where both limits can be achieved.  The linear distortion is then
\begin{equation}
	D_w = \frac{|\partial_z w| + |\partial_{\overline{z}} w|}{|\partial_z w| - |\partial_{\overline{z}} w|} = \frac{1 + |\mu|}{1 - |\mu|}
    \label{qc6}
\end{equation}
where we have defined the {\it Beltrami coefficient} or {\it complex dilatation} $\mu = \partial_{\overline{z}} w / \partial_z w$.  In terms of the Beltrami coefficient, the Jacobian determinant is $J = |\partial_z w|^2 \left( 1-|\mu|^2\right)$.  Therefore, the strict positivity of $J$ required for a sense-preserving diffeomorphism implies that $|\mu| < 1$.  As previously mentioned, a quasiconformal transformation is a mapping of bounded distortion.  In particular, a map is called $K$-quasiconformal if $D_w \leq K$.  It follows that a conformal transformation can be considered as a 1-quasiconformal map.

The local distortion of conformality induced by a quasiconformal map is entirely encoded in the parameter $\mu$.  If we consider an infinitesimal circle at centered a point $z_0$, its image ellipse has a major axis of length $1 + |\mu|$ and a minor axis of length $1-|\mu|$.  Furthermore, the angle of maximal distortion in the $z$-plane is $\alpha = \text{ arg}(\mu)/2$ and the angle of minimal distortion is $\alpha + \pi/2$.  In other words, $\mu$ measures regions that do not preserve local geometry.  This geometric intuition, displayed graphically in the inset of \Fig{fig:growth_and_morphospace}(a), motivates the use of $\mu$ as a measure of the anisotropy associated with the deformation of an elastic body.

We complement this geometric derivation by stating the analytic definition of a quasiconformal map.  A map $f$ is considered quasiconformal if it is a solution to the {\it complex Beltrami equation}
\begin{equation}
	\frac{\partial f}{\partial \overline{z}} = \mu(z, \overline{z}) \frac{\partial f}{\partial z}
    \label{BeltramiEq}
\end{equation}
given a Lebesgue-measurable function $\mu$ with $|\mu|_{\infty} < 1$.  This latter constraint is sufficient to ensure that $J > 0$ everywhere and hence, by the inverse function theorem, that $f$ is a sense-preserving diffeomorphism.  Notice that when $\mu = 0$, the Beltrami equation reduces to the Cauchy-Riemann equation, $\partial_{\overline{z}} f = 0$, the necessary and sufficient conditions for the analyticity of a complex map \cite{needham1998visual}.

Confining our attention to topological disks, the Measurable Riemann Mapping Theorem assures us that, for each admissible Beltrami coefficient $\mu$ defined over $\mathbb{D}$, there is a unique solution $f:\mathbb{D} \rightarrow \mathbb{D}$ to \Eq{BeltramiEq} that fixes the points $f(0) = 0$ and $f(1) = 1$ \cite{Astala2009}. To explore the consequences of this theorem, consider two disk-like surfaces $\mathcal{M}_1, \mathcal{M}_2 \subset \mathbb{R}^3$ . We wish to construct an arbitrary sense-preserving diffeomorphism between them, $\vec{\bf \Phi} : \mathcal{M}_1 \rightarrow \mathcal{M}_2$ that specifies the transformation of the points $\vec{\bf X}_0 \in \mathcal{M}_1$ and $\vec{\bf X}_1 \in \partial \mathcal{M}_1$, i.e. $\vec{\bf X}'_0 = \vec{\bf \Phi}(\vec{\bf X}_0) \in \mathcal{M}_2$ and $\vec{\bf X}'_1 = \vec{\bf \Phi}(\vec{\bf X}_1) \in \partial \mathcal{M}_2$. The celebrated Riemann Mapping Theorem assures us that is is always possible to conformally map all such regions to the unit disk and that such a mapping is unique up to a M\"{o}bius transformation  $\varphi: \mathbb{D} \rightarrow \mathbb{D}$ of the form \cite{needham1998visual}
\begin{equation}
	\varphi(z) = e^{i \theta} \frac{z - z_0}{1 - \overline{z_0} z}, \quad |z_0| < 1, \quad \theta \in [0, 2\pi).
    \label{MobiusDisk}
\end{equation}
Given a pair of such conformal maps, $\vec{\bf R}_1 : \mathbb{D} \rightarrow \mathcal{M}_1$ and $\vec{\bf R}_2 : \mathbb{D} \rightarrow \mathcal{M}_2$ , we can fix this conformal gauge freedom by demanding that $\vec{\bf R}^{-1}_1(\vec{\bf X}_0) = \vec{\bf R}_2^{-1}(\vec{\bf X}_0') = 0$ and  $\vec{\bf R}^{-1}_1(\vec{\bf X}_1) = \vec{\bf R}_2^{-1}(\vec{\bf X}_1') = 1$. In this case, both $\vec{\bf R}_1$ and $\vec{\bf R}_2$ are guaranteed to be unique. The most general admissible mapping $\vec{\bf \Phi} : \mathcal{M}_1 \rightarrow \mathcal{M}_2$ can now be decomposed into $\vec{\bf \Phi} = \vec{\bf R}_2 \circ f \circ \vec{\bf R}_1^{-1}$, where $f : \mathbb{D} \rightarrow \mathbb{D}$ is the unique quasiconformal mapping fixing $f(0) = 1$ and $f(1)= 1$ with associated Beltrami coefficient $\mu$. We therefore arrive at the following relation regarding the space of all diffeomorphisms of topological disks and the space of all Beltrami coefficients
\begin{equation}
    \begin{split}
	    &\frac{\text{Space of Diffeomorphisms}}{\text{M\"{o}bius Transformations}} \\
	    &\qquad \qquad \qquad \cong \text{ Space of Beltrami Coefficients}
	\end{split}
	\label{DiffBeltrami}
\end{equation}
where the congruence symbol denotes an isomorphism.
We pause to emphasize the importance of this observation.  Any arbitrary sense-preserving diffeomorphism connecting two topological disks is a quasiconformal transformation.  Furthermore, the Beltrami coefficient characterizing this transformation uniquely describes any and all local anisotropy associated with the mapping. Finally, given a Beltrami coefficient, $\mu$, and the correspondence between two points, this transformation is also unique.

\subsection{Beltrami Holomorphic Flow}
\label{app:qc:BHF}

The Beltrami Holomorphic Flow (BHF) gives a precise form for the variation of a quasiconformal mapping due to the variation of its associated Beltrami coefficient under suitable normalization. The BHF on the unit disk, $\mathbb{D}$, was deduced by Lui \emph{et al}. in \cite{lui2012bhf}, where it was proposed as a tool for the optimization of registrations between static pairs of surfaces. A more formal treatment of the subject matter can be found therein. For our purposes it is sufficient to consider a time dependent Beltrami coefficient $\mu(z,t)$ defined over $\mathbb{D}$. As $\mu$ changes, it will induce a corresponding flow in the unique quasiconformal map $w$, where $w(0) = 0$, $w(1) = 1$, and $\mu = \partial_{\bar{z}} w / \partial_z w$. This flow is given by
\begin{equation}
	w(z,t+\delta t) = w(z, t) + \delta t \, h_t[w,\dot{\mu}](z) + \mathcal{O}(\delta t^2)
	\label{bhf_flow_1}
\end{equation}
where
\begin{equation}
	h_t[w, \dot{\mu}](z) = \int_{\mathbb{D}} K(z, \zeta) \, d\eta^1 \, d\eta^2,
	\label{og:htmu2}
\end{equation}
and
\begin{equation}
	\begin{split}
		&K(z,\zeta) = -\frac{w(z)\left( w(z) -1 \right)}{\pi} \quad \times\\
		&\quad  \bigg( \frac{\dot{\mu}(\zeta) \left( \partial_{\zeta} w(\zeta) \right)^2 }{w(\zeta) \left( w(\zeta)-1 \right)\left( w(\zeta) - w(z) \right) } \\
		&\quad + \frac{\overline{\dot{\mu}(\zeta)}\,  \overline{\left(\partial_{\zeta} w(\zeta) \right)}^2 }{\overline{w(\zeta)} \left( 1-\overline{w(\zeta)} \right)\left( 1-\overline{w(\zeta)} \, w(z) \right) } \bigg).
	\end{split}
	\label{og:Kmu1}
\end{equation}
for $\zeta = \eta^1 + i\eta^2$.
The following form of $h_t[w, \dot{\mu}](z)$ will be useful when we discuss the discretization of the BHF:
\begin{equation}
	h_t[w,\dot{\mu}](z) = \int_{\mathbb{D}} \, 
	\begin{pmatrix}
		G_1 \, \dot{\mu}_1 + G_2 \, \dot{\mu}_2 \\
		G_3 \, \dot{\mu}_1 + G_4 \, \dot{\mu}_2
	\end{pmatrix}
	\, d\eta^1 \, d\eta^2
	\label{og:Vmu3}
\end{equation}
where $\dot{\mu} = \dot{\mu}_1 + i \, \dot{\mu}_2$ and $G_1, G_2, G_3, G_4$ are real valued
functions defined on $\mathbb{D}$ whose form can be deduced from $K(z,\zeta)$.  Here, we identify $A + i B$ as the column vector
$\begin{pmatrix} A \\ B \end{pmatrix}$. 

The BHF can be exploited to help optimize arbitrary functions of surface diffeomorphisms on topological disks. In particular, it provides us with the appropriate descent directions needed for gradient based optimization methods. Consider some functional
\begin{equation}
    \mathcal{S}[\mu] = \int_{\mathbb{D}} \mathcal{L}[\mu, w] \, dx^1 \, dx^2,
    \label{app:og:func1}
\end{equation}
which may depend on $\mu$ both explicitly and implicitly through the quasiconformal mapping $w$. Direct calculation yields
\begin{equation}
    \begin{split}
    \frac{\delta \mathcal{S}[\mu]}{\delta \mu}(\zeta) &= (\nabla_{\mu} \mathcal{L}[\mu, w])(z)\\
     &+ \int_{\mathbb{D}} \, 
	\begin{pmatrix}
		A(z) \, G_1(z,\zeta) + B(z) \, G_3(z,\zeta) \\
		A(z) \, G_2(z,\zeta) + B(z) \, G_4(z,\zeta)
	\end{pmatrix}
	\, dx^1 \, dx^2,
    \end{split}
    \label{app:og:funcDeriv2}
\end{equation}
where $(\nabla_{\mu} \mathcal{L}[\mu, w])(z) = (\partial \mathcal{L} / \partial \mu_1, \partial \mathcal{L} / \partial \mu_2)^T$, $\begin{pmatrix} A \\ B \end{pmatrix} = (\nabla_{w} \mathcal{L}[\mu, w])(z)$ denotes the gradient of $\mathcal{L}$ with respect to the map $w$, and $\begin{pmatrix} A \\ B \end{pmatrix} \cdot \begin{pmatrix} C \\ D \end{pmatrix}$ should be interpreted as the standard dot product of two vectors. The quantity $\delta \mathcal{S}[\mu]/\delta \mu$ can be used to define a descent direction for a flow on the space of $\mu$ that optimizes $\mathcal{S}[\mu]$.

\appsection{Numerical Methods For Optimal Growth Pattern Selection}
\label{app:num}

Numerical solutions to the problem of optimal growth were generated using a geometric finite difference method \cite{Weischedel2012}. Consider a smooth surface $\mathcal{M}_t \subset \mathbb{R}^3$ parameterized in the usual way by
a set of coordinates $\vec{\bf x} \in \mathbb{R}^2$. In our methods, this smooth parameterized 3D surface was approximated by a mesh triangulation $\mathcal{M} = (\mathcal{F}, \mathcal{V}, \mathcal{E})$, where $\mathcal{V}$ denotes the set of vertices comprising the mesh, $\mathcal{F}$ denotes the connectivity list defining mesh faces, and $\mathcal{E}$ denotes the set of mesh edges, defined parsimoniously through $\mathcal{F}$. Each vertex $v_i \in \mathcal{V}$ carries a set of 3D coordinates, $\vec{\bf R}_i$, corresponding to that vertex's position in physical space, and also a set of 2D coordinates, $\vec{\bf x}_i$, corresponding to that vertex's position in the domain of parameterization. In
order to maintain a fixed mesh topology, we choose to order the vertices
describing each face in a counter-clockwise fashion (\Fig{fig:app_og_num:NES}(a)). The edges of the face are labelled by the vertex index opposite
to that edge. A discrete metric on a triangular mesh is a set of edge lengths, $L : \mathcal{E} \rightarrow \mathbb{R}^+$, which satisfies the triangle inequality on each face $F = [ v_i,v_j, v_k ] \in \mathcal{F}$:
\begin{equation}
    L_i + L_j > L_k, \, L_j + L_k > L_i, \, L_k + L_i > L_j.
    \label{app:num:discMetric1}
\end{equation}
Due to the fact that convex combinations of discrete metrics are still valid discrete metrics, it is possible to show that the space of all discrete Riemannian metrics on a triangular mesh of fixed topology is convex \cite{zeng2013ricci}.

\subsection{Non-Parametric Representations of Discrete Surfaces by Smooth Interpolants}
\label{app:num:NNI}

Computing the optimal growth pattern for a given input requires searching the
space of parameterizations of the final target shape. We must therefore be able
to evaluate dynamical fields on surface configurations corresponding to arbitrary
parameterizations. In order to do so, we store the final surface as a smooth
interpolant over the unit disk $\mathbb{D}$ rather than as an explicit triangulation with fixed topology and 3D vertex positions. In particular, we utilize natural neighbor interpolation for
scattered 2D data points \cite{Sibson1981,Farin1990}.  We will now explain this
method in its general form and then show how it is applied for the specific
case of evaluating surface configurations.

Let $\mathcal{P} = \{ {\bf p}_1, \ldots, {\bf p}_n \}$ be a set of $n$ points
in $\mathbb{R}^2$ and let $\vec{\boldsymbol{\Phi}}$ be a vector-valued function
defined on the convex hull of $\mathcal{P}$.  We assume that the function
values are known at the points of $\mathcal{P}$. In the context of surface
evaluation, these are simply the 3D locations of the data points, i.e.
$\vec{\bf R}_i = \vec{\boldsymbol{\Phi}}({\bf p}_i)$.  The point set
$\mathcal{P}$ defines a Voronoi tessellation of $\mathbb{R}^2$, Equivalently,
there is a unique Delaunay triangulation associated to this point set. We also
require knowledge of the gradient at each point ${\bf G}_i = \nabla
\vec{\boldsymbol{\Phi}}({\bf p}_i)$. If the user has analytic knowledge of the
function $\vec{\boldsymbol{\Phi}}$ these gradients can be supplied as inputs.
Otherwise, they are estimated by fitting high order Taylor polynomials to local
neighborhoods of data points and extracting the first-order coefficients.
Experiments show that fitting a 3rd order Taylor polynomial produces high
quality results without being too computationally expensive.

\begin{figure}[tbhp!]
\centering
\includegraphics[width=0.7\linewidth]{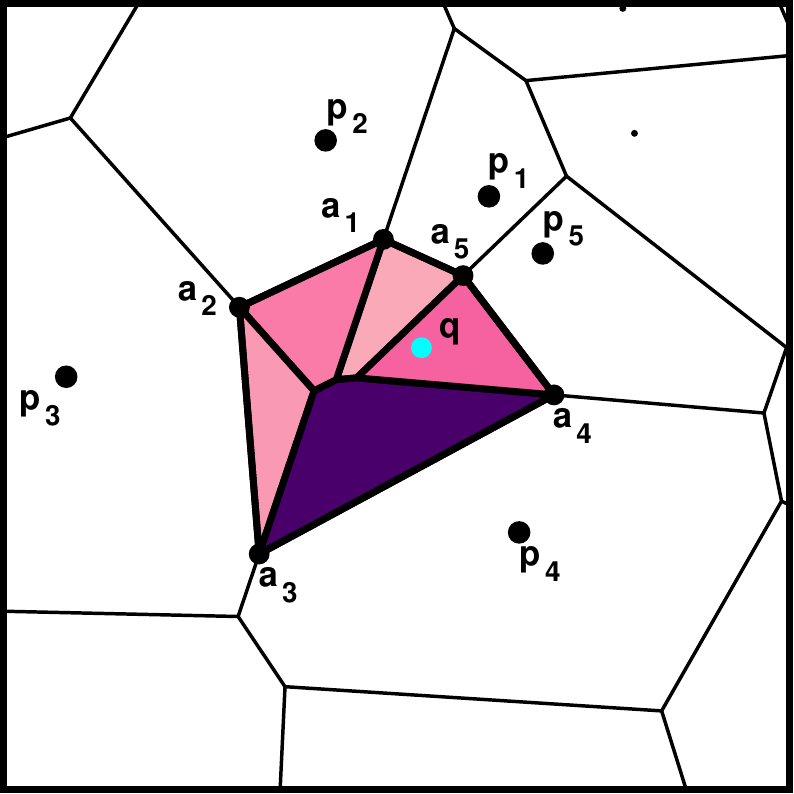}
\caption{Natural neighbor coordinate construction. The set $\{{\bf p}_i\}$ denote the data points whose functional values are being interpolated and the set $\{{\bf a}_i\}$ denote vertices of the virtual Voronoi polygon associated with the query point ${\bf q}$. Colors indicate area weights of data point contributions to the interpolated function value at ${\bf q}$.}
\label{fig:NNICoords}
\end{figure}

The interpolation is carried out for an arbitrary query point ${\bf q}$ on the
convex hull of $\mathcal{P}$. When simulating the insertion of the query point
${\bf q}$ into the Voronoi diagram of $\mathcal{P}$, the virtual Voronoi cell of
${\bf q}$ ``steals'' some area from the existing existing cells.  This
construction is illustrated in Figure \ref{fig:NNICoords}.
Let $A({\bf q})$ denote the area of the virtual Voronoi cell of ${\bf q}$ and
let $A_i({\bf q})$ denote the area of the sub-cell that would be stolen from
the cell of ${\bf p}_i$ by the cell of ${\bf q}$. The natural neighbor
coordinates of ${\bf q}$ with respect to the data point ${\bf p}_i \in
\mathcal{P}$ are defined to be
\begin{equation}
	\lambda_i({\bf q}) = \frac{A_i({\bf q})}{A({\bf q})}.
	\label{NNICoords1}
\end{equation}
These coordinates have the following properties:
\begin{itemize}
	\item ${\bf q} = \sum_{i=1}^n \lambda_i({\bf q}) \, {\bf p}_i$ (barycentric coordinate property)
	\item For any $i,j \leq n$, $\lambda_i({\bf p}_j) = \delta_{ij}$
	\item $\sum_{i=1}^n \lambda_i({\bf q}) = 1$ (partition of unity property)
\end{itemize}
Furthermore, the natural neighbor coordinates depend continuously on the planar
coordinates of ${\bf q} = (q^x, q^y)$. In fact, one can calculate the gradient
of the ``stolen'' sub-cell area $A_i({\bf q})$ with respect to the components
of ${\bf q}$.  Let the $m$ natural neighbors of ${\bf q}$ be denoted $\{ {\bf
p}_1, \ldots, {\bf p}_m \}$ and be arranged in counter-clockwise order around
${\bf q}$ (this numbering is for convenience in the calculation of local
quantities with respect to ${\bf q}$ and does not have to match the global
numbering scheme used in $\mathcal{P}$). Additionally, let the set $\{{\bf a}_1,
\ldots, {\bf a}_m\}$ refer to the counter-clockwise ordered vertices of the
virtual Voronoi cell of ${\bf q}$.  It can be shown that
\begin{equation}
	\nabla A_k({\bf q}) = \frac{f_k}{d_k} \, \left( \frac{{\bf a}_k + {\bf a}_{k+1} }{2} - {\bf q} \right) = \frac{f_k}{d_k} \, \left( \frac{\vec{\bf v}_k + \vec{\bf v}_{k+1}}{2} \right)
	\label{NNIGrad1}
\end{equation}
where $f_k = ||{\bf a}_{k+1} - {\bf a}_k||$, $d_k = || {\bf p}_k - {\bf q} ||$,
and $\vec{\bf v}_k = {\bf a}_k - {\bf q}$. The gradient of $\lambda_i({\bf q})$
follows trivially from application of the chain rule and the fact that $A({\bf
q}) = \sum_k A_k({\bf q})$.

Having calculated the natural neighbor coordinates of ${\bf q}$, we can infer
the function value $\vec{\boldsymbol{\Phi}}({\bf q})$ via interpolation with
respect to these coordinates. In particular, we choose to use Sibson's $C^1$
interpolant \cite{Sibson1981}. Let
\begin{equation}
	\vec{\bf Z}^0({\bf q}) = \sum_i \lambda_i({\bf q}) \, \vec{\boldsymbol{\Phi}}({\bf p}_i)
	\label{Z0}
\end{equation}
denote the linear combination of the neighbor's function values weighted by the natural neighbor coordinates.  Furthermore, we define the functions
\begin{equation}
	\begin{split}
		&\vec{\boldsymbol{\xi}}_i({\bf q}) = \vec{\boldsymbol{\Phi}}({\bf p}_i) + {\bf G}^T_i({\bf q} - {\bf p}_i) \\
		&\qquad \implies \vec{\boldsymbol{\xi}}({\bf q}) = \frac{\sum_i \frac{\lambda_i({\bf q})}{||{\bf q}-{\bf p}_i|| } \, \vec{\bf \xi}_i({\bf q}) }{ \frac{\lambda_i({\bf q})}{||{\bf q}-{\bf p}_i|| } },
	\end{split}
	\label{Z1Xi}
\end{equation}
\begin{equation}
	\alpha({\bf q}) = \frac{\sum_i \lambda_i({\bf q}) \, ||{\bf q}-{\bf p}_i||  }{ \frac{\lambda_i({\bf q})}{||{\bf q}-{\bf p}_i|| } },
	\label{Z1Alpha}
\end{equation}
and
\begin{equation}
	\beta({\bf q}) = \sum_i \lambda_i({\bf q}) \, ||{\bf q} - {\bf p}_i||^2.
	\label{Z1Beta}
\end{equation}
In terms of these quantities, the final interpolant is defined to be
\begin{equation}
	\vec{\bf Z}^1({\bf q}) = \frac{\alpha({\bf q}) \, \vec{\bf Z}^0({\bf q}) + \beta({\bf q}) \, \vec{\boldsymbol{\xi}}({\bf q})}{\alpha({\bf q}) + \beta({\bf q})}
	\label{Z1}
\end{equation}
Sibson noticed that this interpolant is $C^1$ continuous with respect to the
coordinates of the query point ${\bf q}$.  In other words, we can calculate
analytic gradients of $\vec{\bf Z}^1({\bf q})$ with respect to ${\bf q}$
(making use of the gradient in \Eq{NNIGrad1} in a series of chain
rule calculations).  This property makes $\vec{\bf Z}^1$ suitable for use in
gradient based optimization procedures.

\begin{figure}[tbhp!]
\centering
\includegraphics[width=\linewidth]{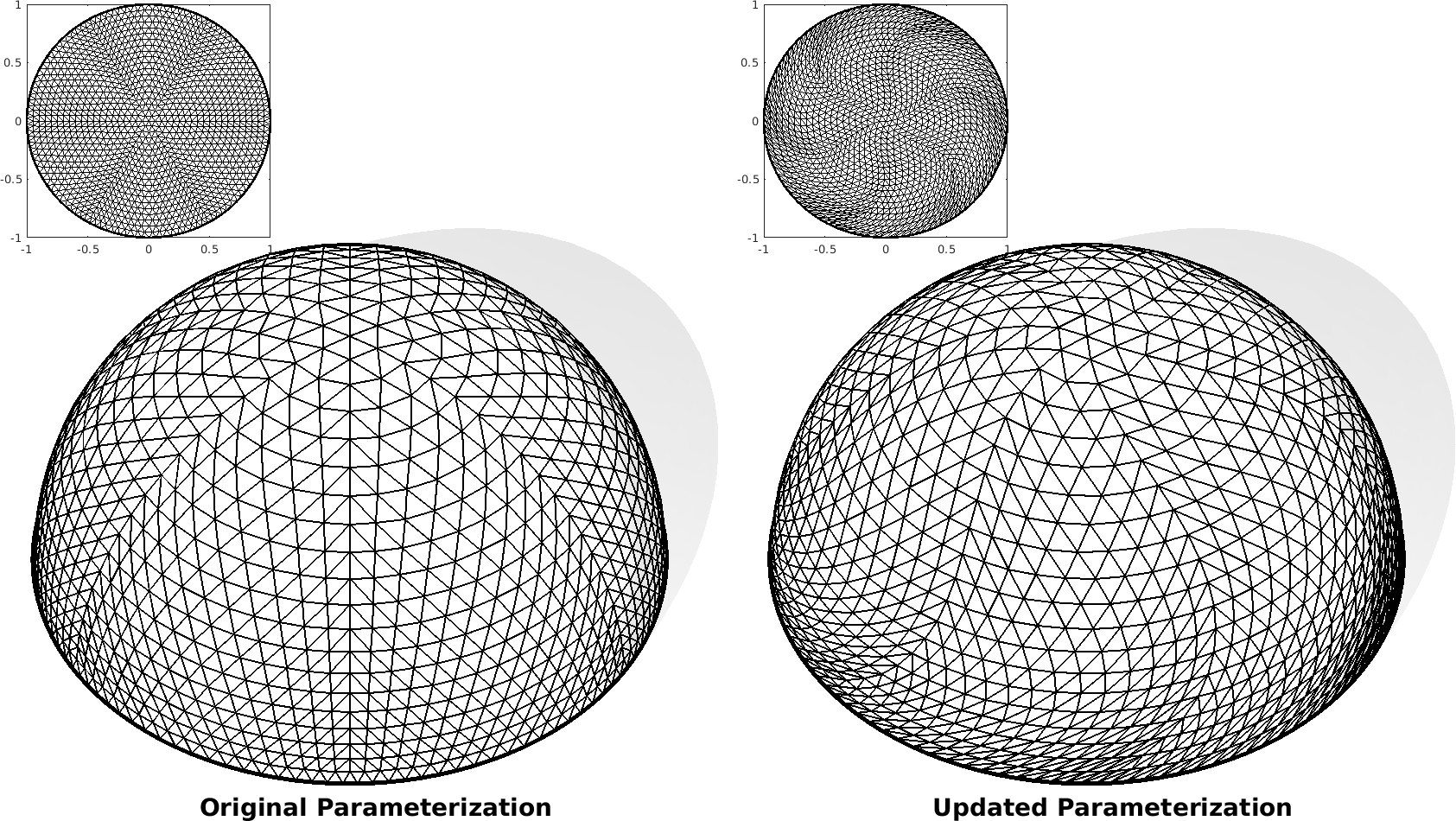}
\caption{Evaluation of surfaces via natural neighbor interpolation for different 2D parameterizations}
\label{fig:NNIParam}
\end{figure}

The application of natural neighbor interpolation to storing surfaces is
straight forward.  We provide as input a set of points $\mathcal{V}$ defining a surface, along with a set of 3D coordinates $\{\vec{\bf R}_i\}$ and 2D coordinates $\{\vec{\bf x}_i\}$ for each point $v_i \in \mathcal{V}$.  Note that we do not need to supply a face
connectivity list $\mathcal{F}$ - any association between the points will be defined through the Delaunay triangulation of $\mathcal{V}$ in $\mathbb{R}^2$. We can now evaluate updated
surface triangulations for {\it any} updated parameterization $\{\vec{\bf x}_i'\}$. This is
shown in Figure \ref{fig:NNIParam}.

\subsection{Discretization of the Beltrami Holopmorphic Flow}

In this section, we explain our numerical computation of the variation in a
quasiconformal map $w:\mathbb{D} \rightarrow \mathbb{D}$ under the variation of
its associated Beltrami coefficient $\mu$. This implementation is based on the
one formulated by Lui \emph{et al}. in \cite{lui2012bhf}. Consider a triangulation of the unit disk
defined by a face connectivity list $\mathcal{F}$ and a set of vertices $\mathcal{V}$,
where $\vec{\bf x}_i \in \mathbb{D}$ denotes the 2D coordinates of the $i$th
vertex $v_i \in \mathcal{V}$. Let $\vec{\bf u}_i = (u^1, u^2) \in \mathbb{D}$ denote the updated
coordinates of the $v_i$ as the result of a quasiconformal mapping.
In the continuous setting, the map $w:\mathbb{D} \rightarrow \mathbb{D}$ has an
associated Beltrami coefficient
\begin{equation}
	\mu = \frac{\partial_{\bar{z}} w}{\partial_z w},
	\label{app:og:disc_mu1}
\end{equation}
where $z = x^1 + i x^2$ and $w = u^1 + i u^2$. In the discrete setting, the map
$w:\mathbb{D} \rightarrow \mathbb{D}$, i.e. updated vertex coordinates, defines
a set of piece-wise constant affine transformations for each triangle in the
mesh. We can define a corresponding piece-wise constant Beltrami coefficient on each face $F \in \mathcal{F}$ by discretizing Eq
\eqref{app:og:disc_mu1} using the finite element method (FEM) gradient
\cite{botsch2010polygon}. For a face $F = [\vec{\bf x}_i, \vec{\bf x}_j,
\vec{\bf x}_k]$, the gradient of a quantity $f$ defined on each vertex is given
by:
\begin{equation}
    \begin{split}
	    &\nabla f = 
	    \left(\partial_x f, \,\partial_y f \right)^T\\
	    &= \frac{1}{2 A_F} \, \left( (f_j-f_i) (\vec{\bf x}_i-\vec{\bf x}_k)^{\perp} + (f_k-f_i) (\vec{\bf x}_j-\vec{\bf x}_i)^{\perp} \right),
	\end{split}
	\label{app:og:fem_grad}
\end{equation}
where $A_F$ is the area of face $F$ and the symbol $\perp$ denotes a
counter-clockwise rotation by $90^{\degree}$ in the plane of the face.
Geometrically, this operation converts a scalar quantity defined on mesh
vertices to a tangent vector defined on each mesh face. The Beltrami cofficient
on each face is therefore given by
\begin{equation}
	\mu_F = \frac{(\partial_{x^1} u^1 - \partial_{x^2} u^2) + i (\partial_{x^1} u^2 + \partial_{x^2} u^1)}{(\partial_{x^1} u^1 + \partial_{x^2} u^2) + i (\partial_{x^1} u^2 - \partial_{x^2} u^1) },
	\label{app:og:disc_mu2}
\end{equation}
where the discrete partial derivative operators are defined through \Eq{app:og:fem_grad}.
For applications where we need to define the Beltrami coefficient on vertices,
we can simply average the face-based quantity in \Eq{app:og:disc_mu2}
\begin{equation}
	\mu_v = \displaystyle\sum_{F \in \mathcal{N}_v} \alpha_F \,\mu_F,
	\label{app:og:disc_mu3}
\end{equation}
using a set of normalized weights $\{\alpha_F\}$, where $\mathcal{N}_v$ denotes the set of faces $F$ attached to vertex $v$. Experiments show that weighting
this average by the normalized internal angle adjacent to the vertex within
each face produces better results than simple averaging or area-weights. For
notational convenience, we define a modified partial derivative operator $D$
that includes this angle weighted averaging step, i.e. $D_{x} f$ maps scalar
vertex quantities to vectors in the tangent space of each vertex. We can define
the vertex-based Beltrami coefficient directly in terms of these modified
operators
\begin{equation}
	\mu_v = \frac{(D_{x^1} u^1 - D_{x^2} u^2) + i \,(D_{x^1} u^2 + D_{x^2} u^1)}{(D_{x^1} u^1 + D_{x^2} u^2) + i \, (D_{x^1} u^2 - D_{x^2} u^1) }.
	\label{app:og:disc_mu4}
\end{equation}

For both quasiconformal map reconstruction and surface function optimization,
the key step is the computation of the variation $h_{\epsilon}[w, \delta \mu]$ of the map
$w$ under the variation $\mu \mapsto \mu + \epsilon \, \delta \mu$. We use $\epsilon$ rather than $t$ and $\delta \mu$ rather than $\dot{\mu}$ to emphasize that the flow optimizing our surface function does not correspond to a physical anisotropy field changing in time. We have
\begin{equation}
	h_{\epsilon}[w, \delta \mu](z) = \int_{\mathbb{D}} K(z, \zeta) \, d\eta^1 \, d\eta^2,
	\label{app:num:Vmu2}
\end{equation}
which is defined explicitly in \Eq{og:Kmu1}, replacing $\dot{\mu}$ with $\delta \mu$. The quantities $w$ and $\delta\mu$ are defined on each vertex $v$. The
derivative $\partial_z w$ can be approximated as
\begin{equation}
	(\partial_z w)_v \approx \frac{(D_{x^1} u^1 + D_{x^2} u^2) + i \,(D_{x^1} u^2 - D_{x^2} u^1) }{2}.
	\label{app:og:dfz_1}
\end{equation}
For each pair of vertices $(v_j, v_k)$, the kernel $K(v_j, v_k)$ can be
assembled on a per-vertex basis in terms of the sets $\{w_v\}$, $\{(\partial_z w)_v\}$, and $\{\delta\mu_v\}$. When $K(v_j, v_k)$ is singular, we set
$K(v_j, v_k) = 0$. Let $A_v$ denote the barycentric area of each vertex, i.e.
\begin{equation}
	A_v = \frac{1}{3} \, \displaystyle\sum_{F \in \mathcal{N}_v} A_F.
	\label{app:og:varea}
\end{equation}
Then, $h_{\epsilon}[w, \delta \mu]$ is can be approximated by
\begin{equation}
	h_{\epsilon}[w, \delta \mu](v_k) = \sum_{v_j} K(v_k, v_j) \, A_{v_j}.
	\label{app:og:disc_bhf1}
\end{equation}
Just as in the continuous setting, we can write out this sum explicitly in terms of its real and imaginary components
\begin{equation}
\begin{split}
	&h_{\epsilon}[w, \delta \mu](v_k) \\
	&= \sum_{v_j} \, 
	\begin{pmatrix}
		G_1(v_k, v_j) \, \delta \mu_1(v_j) + G_2(v_k, v_j) \, \delta \mu_2(v_j) \\
		G_3(v_k, v_j) \, \delta \mu_1(v_j) + G_4(v_k, v_j) \, \delta \mu_2(v_j)
	\end{pmatrix}
	\, A_{v_j},
	\end{split}
	\label{app:og:disc_bhf2}
\end{equation}
where $\delta \mu = \delta \mu_1 + i \,\delta \mu_2$.

\subsection{Evaluation and Minimization of the Optimal Growth Energy}

From \Eq{og:cg_mingro2}, the optimal growth energy for constant growth
patterns with $\dot{\Gamma} = \dot{\gamma} = 0$ is defined to be
\begin{equation} 
    \begin{split}
        \mathcal{S} &= \lambda \, D_{SEM}[{\bf g}_T; \, \mathcal{M}_T] \quad + \\
	    &T \int_{\mathcal{B}} d^2\vec{\bf x}\, \sqrt{g_0} \, \, \left[  c_1 |\nabla {\Gamma}|^2 + c_2|\nabla \gamma|^2 \right].
    \end{split}
    \label{app:og:cg_mingro2}
\end{equation}
In practice, however, implementing such constrained optimization methods is difficult and it is an attractive to reformulate Eq. \eqref{app:og:cg_mingro2} as an unconstrained problem. Such a reformulation is possible by making the approximation $\gamma \approx \dot{\mu}$. From Eq. \eqref{bhf_mu1}, we see that this approximation holds for $|\mu|^2 << 1$ and $e^{i \psi} = (\overline{\partial_z w})/(\partial_z w) \approx 1$. Writing $z = r \, e^{i \theta}$, this latter condition is exactly true, for instance, when $w(z,t) = \rho(r,t)\, e^{i \theta}$, which applies to a broad variety of relevant growth patterns, including the growth pattern generating the hemispherical cap shown in \Fig{fig:og:synth_surf}(a-c). In this new approximation scheme, since $\ddot{\mu} \approx \dot{\gamma} = 0$, the Beltrami coefficient as a function of time is simply $\mu(\vec{\bf x}, t) \approx t\, \dot{\mu}(\vec{\bf x}) \approx t \, \gamma({\bf x})$, where we have exploited the fact that we are always free to choose a conformal parameterization for our initial time point. The modified constant growth functional is given by
\begin{equation}
    \begin{split}
        \tilde{\mathcal{S}} &= \int_{\mathcal{B}} d^2\vec{\bf x}\, \sqrt{g_0} \, \, \left[  T \, c_1 |\nabla {\Gamma}|^2 + T^{-1} \, c_2 |\nabla \mu(T)|^2 \right] \\
    \end{split}
    \label{app:og:cg_mingro3}
\end{equation}
Given an initial configuration, both $\Gamma$ and $\mu(T)$ can be found directly from a candidate final configuration without having to calculate any intermediate geometries. In other words, given an initial configuration and final target shape, we can solve an \emph{unconstrained} optimization problem over the space of parameterizations of the final shape using the BHF to find the fields $\Gamma$ and $\mu(T)$ that minimize the reduced cost function in \Eq{app:og:cg_mingro3}.
Since we are directly optimizing over the space of parameterizations of the final shape, the shape constraint is satisfied by construction, which obviates the need for an explicit Lagrange multiplier.
We can then find all of the intermediate metric tensors using the previously discussed properties of constant growth patterns. With these intermediate metrics in hand, we can embed the time-dependent target geometries in $\mathbb{R}^3$ to generate a complete prediction for both the coarse-grained shape and flow of a growing tissue over time.

The input to our numerical optimization method requires a mesh triangulation
with a set of 3D vertex coordinates $\vec{\bf R}_0$ defining the initial 3D configuration, a set of
2D vertex coordinates defining the Lagrangian parameterization $\vec{\bf x}_0$,
and a set of 3D vertex coordinates $\vec{\bf X}_T$, which will be
converted into a non-parametric representation of the final shape using natural
neighbor interpolation. At each step of the minimization process, the the energy is evaluated in terms of an updated set of 2D coordinates $\vec{\bf w}_T$ and a corresponding set of final 3D vertex coordinates $\vec{\bf R}_T$, for which $\vec{\bf x}_0$ and $\vec{\bf X}_T$ serve as an initial guess. Let $A_{F}^{(0)}$ denote the 3D area of face $F$ in the
initial configuration at time $t = 0$ and let $A_F$ denote the 3D area of face $F$ in the final
configuration at time $t = T$. The constant growth rate $\Gamma$ can be approximated on
each face through
\begin{equation}
	A_F = A_F^{(0)} \, e^{\Gamma_F \, T} \implies \Gamma_F = \frac{1}{T} \, \text{log}\left[\frac{A_F}{A_F^{(0)}}\right],
	\label{app:og:disc_gamma1}
\end{equation}
and then mapped to values on vertices via the same angle-weighted averaging
procedure defined in \Eq{app:og:disc_mu3} for face-based Beltrami
coefficients. The optimal growth energy can therefore be approximated as
\begin{equation}
    \begin{split}
	    \tilde{\mathcal{S}} &= \displaystyle\sum_{F \in \mathcal{F}} A_F^{(0)} \, \left( \tilde{c}_1 \bigg|\bigg| \frac{A_F^{(0)}}{A_F} \, \left(\nabla \frac{A}{A^{(0)}}\right)_F \bigg|\bigg|^2 + \tilde{c}_2 |(\nabla \mu)_F|^2 \right) \\
	    &= \underbrace{\displaystyle\sum_{F \in \mathcal{F}} A_F^{(0)} \, \mathcal{H}(\vec{\bf R}(w[\mu]))_F}_{\equiv \tilde{\mathcal{S}}_1} + \underbrace{\tilde{c}_2 \displaystyle\sum_{F \in \mathcal{F}} A_F^{(0)} \,  |(\nabla \mu)_F|^2}_{\equiv \tilde{\mathcal{S}}_2},
	\end{split}
	\label{app:og:disc_cg_mingro}
\end{equation}
where we have absorbed factors of $1/T$ into the modified constants $\tilde{c}_1$ and $\tilde{c}_2$ for notational simplicity.
In the second equality, we introduced the functional $\mathcal{H}(\vec{\bf R}(w[\mu]))$ to explicitly show that $\tilde{\mathcal{S}}_1$ only depends on $\mu$ through the 3D embedding of the updated planar coordinates $\vec{\bf R}(w[\mu])$. This energy can now be minimized using standard gradient descent methods over the space of vertex based Beltrami coefficients $\mu_v$. 

For our purposes, the gradients of \Eq{app:og:disc_cg_mingro} were calculated by hand. The gradients of $\tilde{\mathcal{S}}_2$, while tedious to tabulate, are conceptually trivial. Calculating the gradients of $\tilde{\mathcal{S}}_1$ requires synthesizing all of the previously described methods comprising our numerical machinery.
\begin{equation}
    \begin{split}
        &\nabla_{\mu} \tilde{\mathcal{S}}_1 = \displaystyle\sum_{F \in \mathcal{F}} A_F^{(0)} \, \nabla_{\mu} \mathcal{H}(\vec{\bf R}(w[\mu]))_F \\
        &= \displaystyle\sum_{F \in \mathcal{F}} A_F^{(0)} \,\left(\nabla_{\vec{\bf R}} \mathcal{H}(\vec{\bf R})_F\right) \cdot \left(\nabla_{w} \vec{\bf R}\right) \cdot \left(\nabla_{\mu} w\right)
    \end{split}
    \label{app:og:muGrad1}
\end{equation}
The gradients $\nabla_{\vec{\bf R}} \mathcal{H}(\vec{\bf R})$ with respect to the 3D vertex coordinates $\vec{\bf R}$ are calculated in the standard way. The gradients $\nabla_{w} \vec{\bf R}$ with respect to the updated 2D vertex coordinates are determined analytically by our natural neighbor interpolation scheme, as detailed in \Appendix{app:num:NNI}. Next, the gradients $\nabla_{\mu} w$ with respect to the vertex-based Beltrami coefficients are calculated according to the BHF scheme detailed in \Appendix{app:qc:BHF}. Finally, the full gradients of $\tilde{\mathcal{S}}_1$ with respect to the $\mu_v$ are computed via the chain rule in terms of these quantities (\Eq{app:og:muGrad1}). For future applications involving some novel optimality criteria $\mathcal{S}'(\vec{\bf R}(w[\mu]))$, the gradients $\nabla_{w} \mathcal{S}'(\vec{\bf R}(w))$ may be computed via automatic differentiation methods optimized for discrete geometry processing , e.g. \cite{schmidt2022tinyad}, and strung together with the application independent gradients $\nabla_{\mu} w$ determined by the BHF. In this way, the useful analytic properties of the BHF can be fully exploited while avoiding the laborious task of manual differentiation.

\subsection{Embedding Intrinsic Geometries in 3D}
\label{app:num:elastic}

Once our numerical machinery has produced a solution for the optimal time course of intrinsic geometry, this geometry must be embedded in $\mathbb{R}^3$ in order to produce a complete prediction for the shape and flow of the growing tissue. As before, we consider a smooth surface $\mathcal{M}_t \subset \mathbb{R}^3$ parameterized in the usual way by
a set of coordinates $\vec{\bf x} \in \mathbb{R}^2$. At each point $\vec{\bf R}
\in \mathcal{M}_t$, The geometry of the surface is captured by the first fundamental
form, $g_{\alpha \beta} = \partial_{\alpha} \vec{\bf R} \cdot \partial_{\beta}
\vec{\bf R}$, and the second fundamental form, $b_{\alpha \beta} =
\partial_{\alpha} \partial_{\beta} \vec{\bf R} \cdot \hat{\bf n}$,  where
$\hat{\bf n}$ is the unit normal to the surface at $\vec{\bf R}$. It is now necessary to explicitly distinguish the physical geometry from the intrinsic geometry.
We characterize the time-dependent intrinsic geometry of the surface in terms the
target metric, $\overline{\bf g}(t)$, and the target curvature tensor,
$\overline{\bf b}(t)$. We define the components of the inverse target metric
tensor by $\bar{g}^{\alpha \beta}$  by $\bar{g}^{\alpha \sigma} \bar{g}_{\sigma
\beta} = \delta^{\alpha}_{\beta}$, so that indices of tensorial quantities are
raised and lowered with respect to the target metric.

We embed the target geometry in $\mathbb{R}^3$ utilizing a method motivated by continuum mechanics. We assume that the tissue is an incompatible
elastic shell with energy
\begin{equation}
	\begin{split}
		&E = E_S + E_B = \frac{Y}{2(1-\nu^2)} \int d\vec{\bf x}^2 \sqrt{\bar{g}}\, \bigg\{ \\ &\frac{h}{4} \left[\, \nu \, \text{Tr}[ \bar{\bf g}^{-1} ( {\bf g} - \bar{\bf g} ) ]^2 + (1-\nu) \, \text{Tr}[(\bar{\bf g}^{-1} ( {\bf g} - \bar{\bf g} ))^2]\, \right] +\\ &\frac{h^3}{12} \, \left[\, \nu \, \text{Tr}[ \bar{\bf g}^{-1} ( {\bf b }- \overline{\bf b} ) ]^2 + (1-\nu) \, \text{Tr}[(\bar{\bf g}^{-1} ({\bf b}-\overline{\bf b}))^2]\, \right] \bigg\},
	\end{split}
	\label{app:og:NES1}
\end{equation}
where $h$ is the thickness of the tissue, $Y$ is Young's modulus, and $\nu$ is
the Poisson ratio. This mechanical energy has been studied both in the context
of morphogenesis and the design and manipulation of synthetic surface
structures for both monolayer and bilayer configurations \cite{sharon2010mechanics,efrati2009uncnep,VanRees2017}. The stretching energy
$E_S \propto h$ describes the mismatch between the target rest lengths and
angles of the surface and the physical rest lengths and angles. Analogously,
the bending energy $E_B \propto h^3$ describes the mismatch between the target
and physical curvatures. Essentially, the system will adopt an equilibrium
configuration that matches the target geometry as much a possible while balancing the competition between stretching and bending.
Importantly, it is not assumed that $\bar{g}$ and $\bar{b}$ satisfy the
Gauss-Codazzi-Mainardi-Peterson compatibility conditions
\cite{frankel2011geophys}. If this is the case, then there is no attainable 3D
configuration for which the energy vanishes identically. In this situation, the
equilibrium configuration will harbor residual stresses. 

\begin{figure}[htpb!]
\centering
\includegraphics{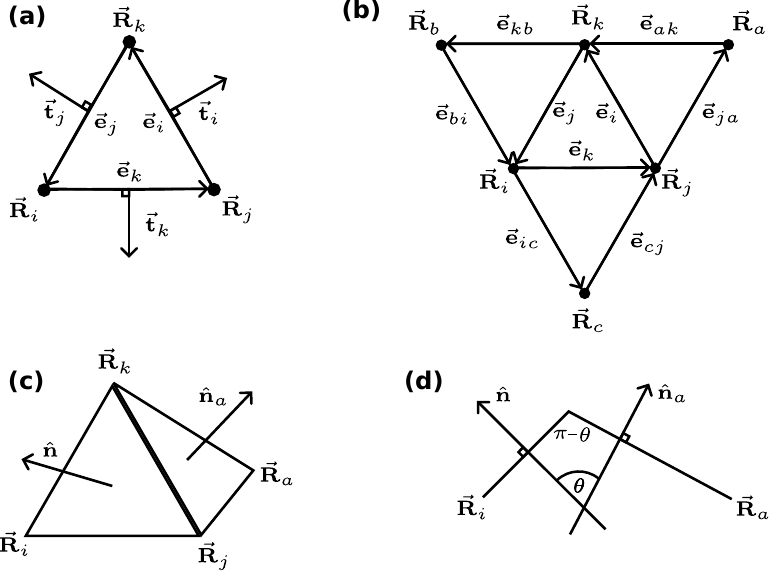}
\caption{Geometry conventions for elastic energy minimization.
(a) Notation for triangle vectors.
(b) The `face-with-flaps' stencil over which the energy is evaluated.
(c) A single hinge stencil viewed from above.
(d) The notation describing the bending angle of a single hinge stencil.}
\label{fig:app_og_num:NES}
\end{figure}

Similarly to the growth pattern optimization machinery, equilibrium configurations of this energy are computed using a geometric finite
difference method \cite{Weischedel2012}. The 3D surface is approximated by a mesh triangulation.  Each face $F \in \mathcal{F}$ is ordered in a counter-clockwise fashion, i.e. $F = [ \vec{\bf R}_i,
\vec{\bf R}_j, \vec{\bf R}_k ]$ where a vertex $\vec{\bf R}_i \in
\mathbb{R}^3$. The edges of the face are labelled by the vertex index opposite
to that edge.  The normal vector of the face, is simply given by $\vec{\bf n} =
\vec{\bf e}_i \times \vec{\bf e}_j = \vec{\bf e}_j \times \vec{\bf e}_k =
\vec{\bf e}_k \times \vec{\bf e}_i$.  The unit normal vector is therefore
$\hat{\bf n} = \vec{\bf n} / 2 A$ where $A$ is the area of the face.  For
simplicity, we define the three in-plane mid-edge normals, $\vec{\bf t}_i$,
$\vec{\bf t}_j$, and $\vec{\bf t}_k$, which are simply the corresponding edge
vectors rotated clockwise by $90^{\degree}$ in the plane of the face, i.e. $\vec{\bf t}_i = \vec{\bf e}_i \times
\hat{\bf n}$.  This construction is illustrated in \Fig{fig:app_og_num:NES}(a).

Since the total elastic energy of the surface is calculated by evaluating an
integral, we must choose a stencil of finite area to serve as the discrete
analog of the integral measure.  For this purpose we choose a
`face-with-flaps' stencil.  The structure and nomenclature associated with
this stencil is shown in \Fig{fig:app_og_num:NES}(b). The extra flaps are
included in order to evaluate the bending energy.  Here, the bending energy is
evaluated at the shared edges of the triangulation, i.e. along the
triangulation hinges.  The structure of a single hinge is shown in
\Fig{fig:app_og_num:NES}(c) and \Fig{fig:app_og_num:NES}(d).  Notice that
$\theta$ denotes the {\it bending} or {\it hinge} angle between the unit normal
vectors of adjacent faces.  The dihedral angle of the edge is therefore $\pi
- \theta$.  In general, $\theta$ is a {\it signed} quantity which depends on
edge orientation.  The sign of $\theta$ for a given edge orientation is chosen
arbitrarily, but consistently, to be the same as the sign of $ \left( \hat{\bf
n}_1 \times \hat{ \bf n}_2 \right) \cdot \vec{\bf e}_0$, where $\hat{\bf n}_1$
is the unit normal of the current face, $\hat{\bf n}_2$ is the unit normal of
the adjacent face flap, and $\vec{\bf e}_0$ is the associated edge with
orientation defined counter-clockwise relative to $\hat{\bf n}_1$.  In this
way, $\theta$ is positive when the normals point away from each other and
negative when they point towards each other. 

Our goal is to build a discrete formulation of the energy in Eq
\eqref{app:og:NES1}. The target geometry of the system can be represented by a
set of target edge lengths and bending angles. The target edge lengths must
satisfy the triangle inequality on each face in order to constitute a valid
geometry (see \Eq{app:num:discMetric1}). In the discrete setting, the tensors ${\bf g}$ and ${\bf b}$ are
approximated by piece-wise constant symmetric $2\times2$ matrices defined on
mesh faces. The components of any such matrix defined over a subset of
$\mathbb{R}^2$ can be uniquely determined by the its action on three linearly
independent vectors in the plane. Analogously to its function in the continuous
setting, the discrete metric tensor on a face is the unique matrix that returns
$\vec{\bf e}_i^T \, {\bf g} \, \vec{\bf e}_i= \ell^2_i$ for each edge $\vec{\bf
e_i}$ in the face with length $\ell_i$. Similarly, the discrete curvature
tensor should encode the bending angles on each edge. In terms of these
representations, we define the discrete strain tensor 
\begin{equation}
	\begin{split}
		\boldsymbol{\varepsilon} & = \bar{\bf g}^{-1} \left( {\bf g} - \bar{ \bf g} \right) \\
		& = -\frac{1}{8 \bar{A}^2} \,\sum_{(ijk)} \left[ \ell_i^2- \ell_j^2 - \ell_k^2  - \left( L_i^2-L_j^2-L_k^2 \right) \right] \, \underline{\vec{\bf t}}_i \otimes \vec{\bf \underline{t}}_i \\
		& = -\frac{1}{8 \bar{A}^2} \sum_{(ijk)} \left[ \varepsilon(\ell_i, L_i) - \varepsilon(\ell_j, L_j) - \varepsilon(\ell_k, L_k) \right] \underline{\vec{\bf t}}_i \otimes \underline{\vec{\bf t}}_i \\
		& = -\frac{1}{8 \bar{A}^2} \, \sum_{(ijk)} E_i \, \underline{\vec{\bf t}}_i \otimes \vec{\bf \underline{t}}_i
	\end{split}
	\label{strain1}
\end{equation}
Here, $\ell_i$ is the length of edge $\vec{\bf e}_i$ in the physical triangulation, $L_i$
is the target length of that edge, $\bar{A}$ is the target area of the face and
$\underline{\vec{\bf t}}$ is the in-plane mid-edge normal of the {\it target}
geometry. The $\bar{A}$ may be calculated directly from the $L_i$ without
reference to an explicit embedding of the target geometry. It may seem
counter-intuitive to define this quantity in terms of the $\underline{\vec{\bf
t}}_i$, since we do not have access to these target quantities directly. We will
see that the final energy is formulated in way that can be calculated without
an explicit embedding, just like the $\bar{A}$.  The tensor product of two
vectors $\vec{\bf u}$ and $\vec{\bf v}$ is computed in a matrix representation
as $\vec{\bf u}_i \otimes \vec{\bf v}_i = \vec{\bf u}_i \, \vec{\bf v}_i^T$.
The notation $(ijk)$ implies a cyclic sum over the edges.  We also define the
function $\varepsilon(\ell_m, L_m) \equiv \ell_m^2 - L_m^2$.  Here, $E_i$ is a
temporary shorthand we use to denote the cyclic sum $\varepsilon_i -
\varepsilon_j - \varepsilon_k$. The discrete bending moment is
\begin{align}
    {\bf B} &= \bar{\bf g}^{-1} \, ( {\bf b} - \overline{\bf b} ) \nonumber\\
    &= \sum_{(ijk)} \frac{1}{2 \bar{A} L_i} \bigg[ \underbrace{2 \tan( \frac{\theta_i}{2} )}_{\varphi(\theta_i)} - \underbrace{ 2 \tan( \frac{\Theta_i}{2} ) }_{\varphi(\Theta_i)}  \bigg] \underline{\vec{ \bf t}}_i \otimes \underline{\vec{\bf t}}_i \\
    &= \sum_{(ijk)} \frac{\Phi(\theta_i, \Theta_i)}{2 \bar{A} L_i} \,  \underline{\vec{ \bf t}}_i \otimes \underline{\vec{\bf t}}_i \nonumber
	\label{shape1}
\end{align}
where $\theta_i$ is the bending angle of edge $\vec{\bf e}_i$ in the physical triangulation, $\Theta_i$ is the target bending angle for that edge and $\Phi(\theta_i)
\equiv \varphi(\theta_i) - \varphi(\Theta_i)$. 

Direct calculation shows that for any symmetric quadratic form ${\bf Q}$
defined on on a face (i.e. a section of $\mathbb{R}^2$) with the following structure
\begin{equation}
	\begin{split}
		{\bf Q} &= \sum_{(ijk)} Q_i  \, \underline{\vec{ \bf t}}_i \otimes \underline{\vec{\bf t}}_i \\
		& \implies \text{Tr}[{\bf Q}] = \sum_i Q_i L^2 \\
		& \implies \text{Tr}[{\bf Q}^2] = \sum_i \sum_j Q_i Q_j ( \underline{\vec{\bf e}}_i \cdot \underline{\vec{\bf e}}_j )^2
	\end{split}
	\label{TraceForm}
\end{equation}

Putting all of this together, the (re-scaled) discrete elastic energy of a
Non-Euclidean shell is given by

\begin{widetext}
\begin{equation}
    \begin{split}
		&\tilde{E} = E/h= \frac{1}{2} \sum_T  \bar{A}_T \, \bigg\{ \frac{1}{4} \, \left[ \nu \left( \sum_i \frac{-E_i}{8 \bar{A}_T^2} L_i^2 \right)^2 + (1- \nu) \left( \sum_i \sum_j \frac{ E_i \, E_j }{64 \bar{A}_T^4} ( \underline{\vec{\bf e}}_i \cdot \underline{\vec{\bf e}}_j )^2 \right) \right] \\
		&\qquad \qquad \qquad + \frac{h^2}{12} \left[ \nu \left( \sum_i \frac{\Phi(\theta_i, \Theta_i) L_i}{2 \bar{A}_T} \right)^2 + (1-\nu) \left( \sum_i \sum_j \frac{\Phi(\theta_i, \Theta_i)\, \Phi(\theta_j, \Theta_j)}{4 \bar{A}_T^2 L_i L_j} ( \underline{\vec{\bf e}}_i \cdot \underline{\vec{\bf e}}_j )^2   \right) \right] \bigg\}
	\label{DiscreteEnergy}
    \end{split}
\end{equation}
\end{widetext}
where the sum is over the faces $T$ of the triangulation and we have set $Y =
1-\nu^2$ ($Y$ is irrelevant to the computation of the equilibrium configuration
since it simply rescales the energy).  The terms $ ( \underline{\vec{\bf e}}_i \cdot \underline{\vec{\bf e}}_j )$ can be determined entirely in terms of the target edge lengths. We minimize this energy using a a custom
built quasi-Newton method using an L-BFGS Hessian approximation
\cite{nocedal2006numerical}.

Strictly speaking, the Bonnet theorem tells us that it is necessary to specify both
$\overline{\bf g}(t)$ and $\overline{\bf b}(t)$ to uniquely specify the surface up
to a rigid motion \cite{frankel2011geophys}.  For small $h$, however, the
bending energy is small relative to the stretching energy, i.e. $E_B << E_S$.
In this regime, the system's tendency to match ${\bf g}(t)$ and $\bar{\bf
g}(t)$ will overwhelm it's tendency to minimize the bending energy. The
optimization of Eq. \ref{app:og:NES1} essentially becomes a machinery for
producing isometric embeddings of the target metric $\bar{\bf g}(t)$ with $E_B$
playing the role of a regularizer.

This machinery for computing the embeddings of an instantaneous target geometry
can also be used to generate embeddings of entire time courses of growth
patterns.  When the timescale of growth in the system (i.e. rate of cell
division, etc.) is long compared to the timescale of mechanical relaxation, the
tissue will always effectively remain in mechanical equilibrium. In this
quasistatic regime, the physical configuration will always be a minimizer of
the energy Eq. \eqref{app:og:NES1} given an instantaneous target geometry.
Over a short time $\Delta t$, the target geometry will change, i.e $\bar{\bf
g}(t) \rightarrow \bar{\bf g}(t+\Delta t)$ and $\overline{\bf b}(t) \rightarrow
\overline{\bf b}(t + \Delta t)$. At each new time step, we minimize the new
elastic energy using the previous time point's configuration as an initial
guess. The result is a full time course of growth embedding the various target
geometries. For our purposes, we chose to set $\overline{\bf b}(t)$ by interpolating the angles between the supplied initial configuration and the optimal final configuration. However, numerical experiments setting $\overline{\bf b}(t)=0$, i.e. a plate-like target geometry, showed that time-dependent growth pattern results were not strongly sensitive to the choice of $\overline{\bf b}(t)$ for sufficiently small $\Delta t$. Larger steps may be possible, for instance, through the implementation of a regime that calculates updates $\overline{\bf b}(t) \rightarrow
\overline{\bf b}(t + \Delta t)$ that are always compatible with the optimal target metric \cite{Wang2012}.

\appsection{Analysis of Growing Appendage in {\large\it P\lowercase{arhyale hawaiensis}}}
\label{app:parhyale}

The recording of the transgenic {\it Parhyale} embryo with a construct for heat-inducible expression of a nuclear marker (H2B-mRFPruby) was generated using multi-view lightsheet fluorescence microscopy (LSFM) beginning 3 days after egg lay (AEL). More details regarding data acquisition and pre-processing can be found in \cite{Wolff2018}.

Our analysis focused on a period of dramatic outgrowth in the T2 appendage from $95-109$h AEL and utilized tissue cartography methods to generate coarse-grained flow patterns on cells on the growing limb \cite{heemskerk2015imsane,Mitchell2022Tubular}. Segmented nuclei locations from \cite{Wolff2018} were used as seed points to reconstruct the limb surfaces. Sparse nuclei locations were used to generate smoothed, up-sampled point clouds corresponding to the tissue surfaces \cite{Huang2013}. These point clouds were then triangulated using an advancing front surface reconstruction method \cite{Cohen-Steiner2004}. Time dependent surfaces were then mapped conformally into the unit disk using a custom-implementation of the discrete Ricci flow \cite{zeng2013ricci}. The conformal degrees of freedom in the time dependent parameterization were pinned by finding an optimal M\"{o}bius transformation that matched the neighborhood structure of nuclei locations at subsequent times without explicit reference to nuclei identity \cite{Le2016}.

Once pulled back into the plane, an updated tracking for the nuclei was performed in 2D using a custom built MATLAB GUI enabling the reconstruction of nuclear lineages and cell tracks. Subsequent time points were then registered using an algorithm that produces smooth quasi-conformal mappings subject to nontrivial landmark constraints \cite{lui2014qchr}. The Beltrami coefficients from these instantaneous mappings were then smoothed spectrally by decomposing them onto a basis of the eigenvectors of the discrete Laplace-Beltrami operator defined on the 2D triangulations and keeping only the lowest five modes \cite{botsch2010polygon}. The full time course of material flow was then reconstructed by composing the infinitesimal mappings produced from the smooth Beltrami coefficients using our implementation of the BHF for mapping reconstruction \cite{lui2012bhf}.
\vfill
\pagebreak

\bibliography{optimal_growth}

\end{document}